\documentclass[paper]{JHEP3}

\usepackage{graphicx}
\usepackage{amsmath}

\newcommand{\ms}{\mskip 1.5mu}
\newcommand{\bs}{\mskip -1.5mu}

\newcommand{\half}{ {\textstyle\frac{1}{2}} }

\newcommand{\im}{\operatorname{Im}}
\newcommand{\re}{\operatorname{Re}}

\newcommand{\gev}{\operatorname{GeV}}

\newcommand{\tvec}[1]{\boldsymbol{#1}}

\newcommand{\ru}[4]{u^{\ms #1 #2}_{#3 #4}}
\newcommand{\rl}[4]{l^{\ms #1 #2}_{#3 #4}}
\newcommand{\rs}[4]{s^{\ms #1 #2}_{#3 #4}}
\newcommand{\rn}[4]{n^{\ms #1 #2}_{#3 #4}}

\newcommand{\ruu}[4]{\tilde{u}^{\ms #1 #2}_{#3 #4}}
\newcommand{\rnu}[4]{\tilde{n}^{\ms #1 #2}_{#3 #4}}

\newcommand{\cost}{\cos\theta_\gamma}
\newcommand{\sint}{\sin\theta_\gamma}

\newcommand{\0}{{\mskip 2.5mu} 0 {\mskip 2.5mu}}

\preprint{DESY-07-049 \\
arXiv:0704.1565 [hep-ph]}

\title{Vector meson production from a polarized nucleon}

\author{M. Diehl \\
Deutsches Elektronen-Synchroton DESY, 22603 Hamburg, Germany
}

\abstract{We provide a framework to analyze the electroproduction
  process $ep\to ep\rho$ with a polarized target, writing the angular
  distribution of the $\rho$ decay products in terms of spin density
  matrix elements that parameterize the hadronic subprocess $\gamma^*
  p\to \rho\ms p$.  Using the helicity basis for both photon and
  meson, we find a representation in which the expressions for a
  polarized and unpolarized target are related by simple substitution
  rules.}

\keywords{Lepton-Nucleon Scattering, Spin and Polarization Effects}

\begin{document}

\section{Introduction}
\label{sec:intro}

Exclusive vector meson production has long played an important role in
studying the strong interaction.  The seminal work
\cite{Radyushkin:1996ru,Collins:1996fb} has renewed interest in this
process, showing that in Bjorken kinematics it provides access to
generalized parton distributions and thus to a wealth of information
on the structure of the proton.  While most theoretical and
experimental studies so far are for an unpolarized proton, the
particular interest of target polarization became clear when it was
pointed out that meson production on a transversely polarized target
is sensitive to the nucleon helicity-flip distribution $E$
\cite{Goeke:2001tz,Ellinghaus:2005uc}.  This distribution offers
unique views on the orbital angular momentum carried by partons in the
proton~\cite{Ji:1996ek,Burkardt:2005km} and on the correlation between
polarization and the spatial distribution of
partons~\cite{Burkardt:2002hr}.  Whereas the corresponding
polarization asymmetry in deeply virtual Compton scattering is under
better theoretical control, vector meson production has the advantage
of a greater sensitivity to the distribution of gluons (which in
Compton scattering only enters at next-to-leading order in
$\alpha_s$).  This holds not only in the high-energy regime but even
in a wide range of fixed-target kinematics
\cite{Diehl:2004wj,Diehl:2005gn,Goloskokov:2006hr}, where polarization
measurements are feasible at existing or planned experimental
facilities.

A different motivation to study polarized exclusive $\rho$ production
is that this channel plays a rather prominent role in semi-inclusive
pion production
\cite{Uleshchenko:2002ht,Airapetian:2004tw,Diehl:2005gn}, which has
become a privileged tool to study a variety of spin effects, see e.g.\
\cite{Vogelsang:2003eb}.  It is important to identify kinematical
regions where the exclusive channel $ep\to ep\rho \to ep\ms
\pi^+\pi^-$ dominates semi-inclusive observables, because in these
regions great care must be taken when interpreting the data in terms
of semi-inclusive factorization.

Even with an unpolarized target, the spin structure of the process
$ep\to ep\rho \to ep\ms \pi^+\pi^-$ is very rich, because the angular
distribution of the final state contains information on the helicities
of the exchanged virtual photon and of the $\rho$ meson, as was worked
out in the classical analysis of Schilling and Wolf
\cite{Schilling:1973ag}.  Yet more detailed information is available
with target polarization \cite{Fraas:1974vx}.  Experiments on
unpolarized targets have found that $s$-channel helicity is
approximately conserved in the transition from the $\gamma^*$ to the
$\rho$, with helicity changing amplitudes occurring at most at the
$10\%$ level
\cite{Adams:1997bh,Breitweg:1999fm,Adloff:2002tb,HERMES-SDME,COMPASS-SDME}.
This greatly simplifies the spin structure of the process.  The aim of
the present paper is to provide an analysis framework for exclusive
$\rho$ production on a polarized nucleon target, making as explicit as
possible the relation between the angular dependence of the cross
section and the helicity amplitudes describing the hadronic subprocess
$\gamma^* p\to \rho p$.  We will present our results in a form that
emphasizes the close similarity in structure between an unpolarized
and a polarized target.  Using the helicity basis for both virtual
photon and meson, we also provide an alternative to the representation
of the unpolarized cross section in \cite{Schilling:1973ag}.

The following section gives the definitions of the kinematics and
polarization variables for the reaction under study.  In
Section~\ref{sec:sdm} we define the helicity amplitudes and the spin
density matrix elements describing the process and discuss some of
their general properties.  In Section~\ref{sec:cross} we express the
angular distribution of the polarized cross section in terms of these
spin density matrix elements and point out some salient features of
this representation.  The simplifications arising from distinguishing
natural and unnatural parity exchange in the reaction are discussed in
Section~\ref{sec:nat-par}.  A number of positivity bounds relating
different spin density matrix elements are given in
Section~\ref{sec:pos}.  In Section~\ref{sec:mix} we explain the
complications arising from the distinction between target polarization
relative to the momentum of either the incident lepton or the virtual
photon.  The role of non-resonant contributions in $\pi^+\pi^-$
production is briefly discussed in Section~\ref{sec:non}.  Our results
are summarized in Section~\ref{sec:sum}.


\section{Kinematics and target polarization}
\label{sec:kin}

Let us consider the electroproduction process
\begin{equation}
  \label{production}
e(l) + p(p) \to e(l') + p(p') + \rho(q') 
\end{equation}
followed by the decay
\begin{equation}
  \label{decay}
\rho(q') \to \pi^+(k) + \pi^-(k') ,
\end{equation}
where four-momenta are given in parentheses.  Throughout this work we
use the one-photon exchange approximation.  All or results are equally
valid for the production of a $\phi$ followed by the decay $\phi\to
K^+ K^-$.  They also hold if the scattered proton is replaced by an
inclusive system $X$ with four-momentum $p'$, as explained at the end
of Section~\ref{sec:sdm}.

\FIGURE{
\includegraphics[width=0.8\textwidth]{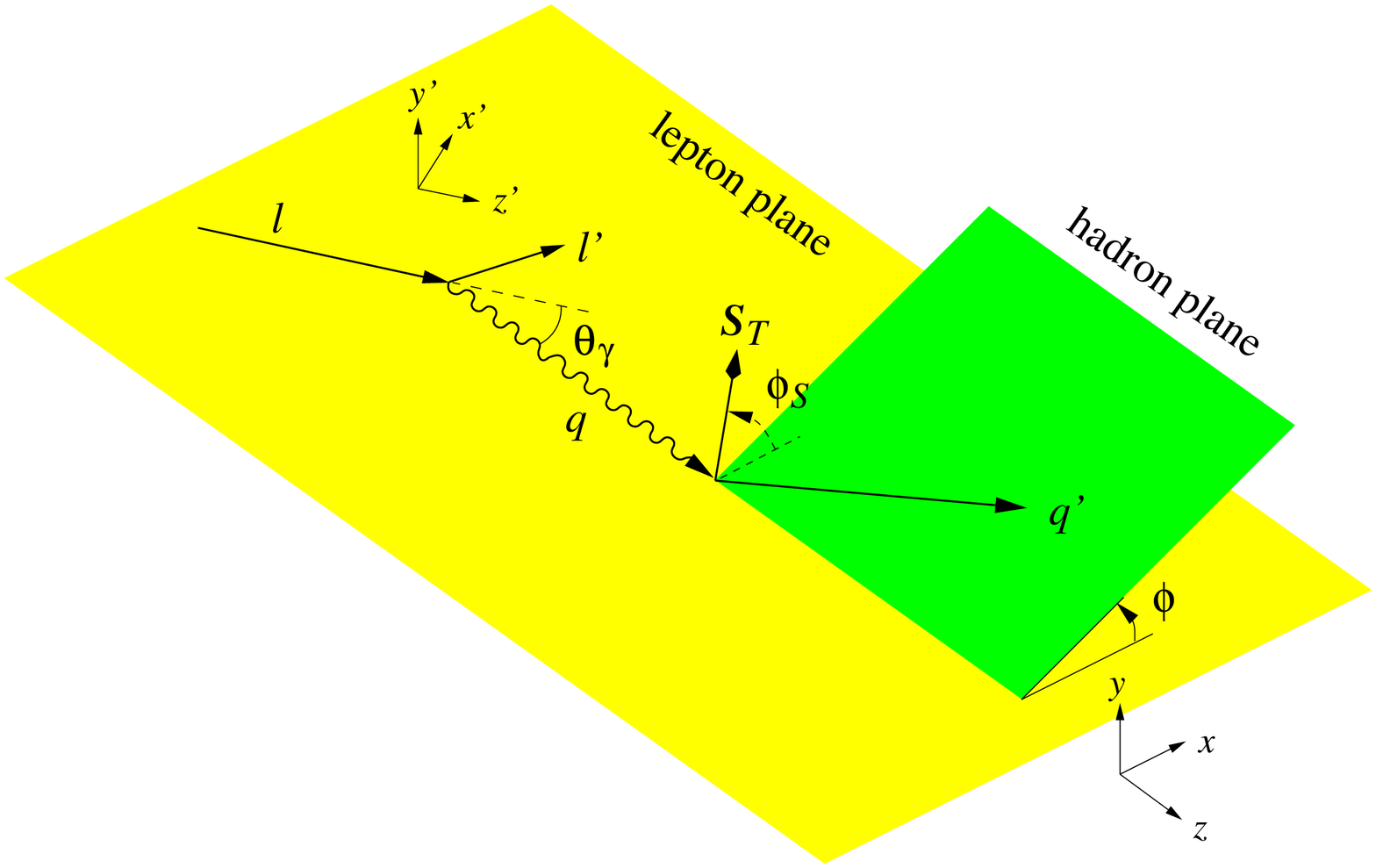}
\caption{\label{fig:trento} Kinematics of $ep\to ep \rho$ in the
  target rest frame.  $\tvec{S}_T$ is the transverse component of the
  target spin vector w.r.t.\ the virtual photon direction.}
}

\FIGURE[b]{
\includegraphics[width=0.95\textwidth]{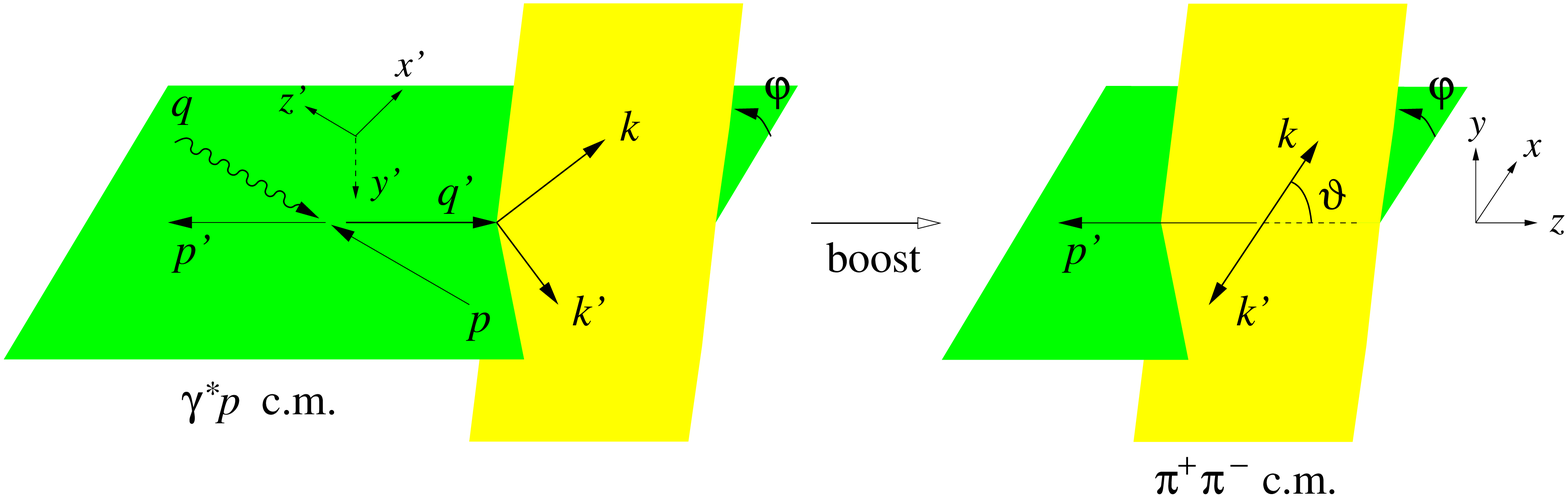}
\caption{\label{fig:rho} Kinematics of the hadronic subprocess
  $\gamma^*p\to \rho\ms p$ followed by the decay $\rho\to \pi^+
  \pi^-$.  The coordinate systems $(x,y,z)$ and $(x',y',z')$ differ
  from those in Fig.~\protect\ref{fig:trento}.}
}

To describe the kinematics we use the conventional variables for deep
inelastic processes, $Q^2= -q^2$, $x_B= Q^2 /(2 p\cdot q)$ and
$y=(p\cdot q) /(p\cdot l)$.  We neglect the lepton mass throughout and
denote the longitudinal lepton beam polarization by $P_\ell$, with
$P_\ell=+1$ corresponding to a purely right-handed and $P_\ell=-1$ to
a purely left-handed beam.
Let us now go to the target rest frame and introduce the right-handed
coordinate system $(x,y,z)$ of Fig.~\ref{fig:trento} such that
$\tvec{q}$ points in the positive $z$ direction and $\tvec{l}$ has a
positive $x$ component.  In this system we have $\tvec{l} =
|\tvec{l}|\ms (\sint, 0, \cost)$ and $\tvec{q} = |\tvec{q}|\ms (0, 0,
1)$, where the angle $\theta_\gamma$ between $\tvec{l}$ and $\tvec{q}$
is defined to be between $0$ and $\pi$.  In accordance with the Trento
convention \cite{Bacchetta:2004jz} we define the angle $\phi$ between
the lepton and the hadron plane as the azimuthal angle of $\tvec{q}'$
in this coordinate system, and $\phi_S$ as the azimuthal angle of the
target spin vector $\tvec{S}$.  Following \cite{Diehl:2005pc} we write
$\tvec{S} = (S_T \cos\phi_S, S_T \sin\phi_S, -S_L)$ with $0\le S_T \le
1$ and $-1 \le S_L \le 1$, so that $S_T$ and $S_L$ describe transverse
and longitudinal polarization with respect to the virtual photon
momentum, with $S_L=1$ corresponding to a right-handed proton in the
$\gamma^* p$~c.m.

To describe the target polarization of a given experimental setup, we
introduce another right-handed coordinate system $(x',y',z')$ in the
target rest frame such that $\tvec{l} = |\tvec{l}|\ms (0,0,1)$ and
$\tvec{q} = |\tvec{q}|\ms (-\sint, 0, \cost)$ as shown in
Fig.~\ref{fig:trento}.  In this system we write $\tvec{S} = (P_T
\cos\psi, P_T \sin\psi, -P_L)$ with $0 \le P_T \le 1$ and $-1 \le P_L
\le 1$, following again \cite{Diehl:2005pc}.  $P_T$ and $P_L$ describe
transverse and longitudinal polarization with respect to the lepton
beam direction, with $P_L=1$ corresponding to a right-handed proton in
the $ep$ c.m.  The two sets of variables describing the target
polarization are related by
\begin{align}
  \label{S-vs-P}
S_T \cos\phi_S &= \cost\ms P_T \cos\psi - \sint\ms P_L \,,
\nonumber \\
S_T \sin\phi_S &= P_T \sin\psi \,,
\nonumber \\
S_L            &= \sint\ms P_T \cos\psi + \cost\ms P_L \,,
\end{align}
which we will use in Sect.~\ref{sec:mix}.  In terms of invariants the
mixing angle $\theta_\gamma$ is given by
\begin{align}
  \label{theta}
\sint &= \gamma\, 
  \sqrt{\frac{1-y- \frac{1}{4} y^2 \gamma^2}{1+\gamma^2}} \,, 
&
\gamma &= \frac{2 x_B\ms M_N}{Q} \,,
\end{align}
where $M_N$ is the nucleon mass.  In Bjorken kinematics $\gamma$ is
small, and so is $\sint \approx \gamma \sqrt{1-y}$.

We finally specify the variables describing the vector meson decay
\eqref{decay}.  This is conveniently done in the $\pi^+\pi^-$ c.m.,
which can be obtained from the $\gamma^* p$ c.m.\ by a boost in the
direction of the scattered nucleon as shown in Fig.~\ref{fig:rho}.  In
the $\pi^+\pi^-$ c.m.\ we introduce the right-handed coordinate system
$(x,y,z)$ shown in Fig.~\ref{fig:rho}, where $\tvec{p}' =
|\tvec{p}'|\ms (0,0,-1)$ and where the target momentum $\tvec{p}$ has
a positive $x$ component.  In this system we define $\vartheta$ and
$\varphi$ as the polar and azimuthal angle of the $\pi^+$ momentum,
i.e.\ $\tvec{k} = |\tvec{k}|\ms (\sin\vartheta \cos\varphi,
\sin\vartheta \sin\varphi, \cos\vartheta)$.  The relation between our
notation here and the one of Schilling and Wolf is\footnote{%
  We remark that the expression for $\sin\Phi$ given in eq.~(13) of
  \protect\cite{Schilling:1973ag} is incorrect since it is always
  positive.  A correct definition is given in
  \protect\cite{Joos:1976nm}.}
\begin{align}
\phi_{\ms\text{here}}^{}
  &= - \Phi_{\text{\protect\cite{Schilling:1973ag}}} \,,
&
\varphi_{\ms\text{here}}^{}
  &= \phi_{\ms\text{\protect\cite{Schilling:1973ag}}} \,,
&
\vartheta_{\ms\text{here}}^{}
  &= \theta_{\ms\text{\protect\cite{Schilling:1973ag}}} \,.
\end{align}


\section{Helicity amplitudes and spin density matrix}
\label{sec:sdm}

The strong-interaction dynamics of the electroproduction process
\eqref{production} is fully contained in the helicity amplitudes for
the subprocess $\gamma^* p\to \rho\ms p$.  From these we will
construct spin density matrix elements which describe the angular
distribution of the overall reaction $ep \to ep\, \pi^+\pi^-$ and its
dependence on the target polarization.

Since we will deal with interference terms we must specify our phase
conventions.  We do this in the $\gamma^* p$ c.m.\ and use the
right-handed coordinate system $(x',y',z')$ shown in
Fig.~\ref{fig:rho}.  In this system we have $\tvec{q} = |\tvec{q}|\ms
(0,0,-1)$ and $\tvec{q}' = |\tvec{q}'|\ms (\sin\Theta, 0,
-\cos\Theta)$, with the scattering angle $\Theta$ of the vector meson
defined to be between $0$ and $\pi$.  Note that the positive $z'$ axis
points along $\tvec{p}$ rather than $\tvec{q}$, as is often preferred
for theoretical calculations.  We specify polarization states of the
target proton by two-component spinors $\chi_{+1/2} = (1,0)$ for
positive and $\chi_{-1/2} = (0,1)$ for negative helicity.  For the
polarization vectors of the virtual photon we choose
\begin{align}
  \label{photon-eps}
\varepsilon_{+1} &= -\frac{1}{\sqrt{2}}\, \bigl( 0,1,-i,0 \bigr) \,,
&
\varepsilon_{-1} &= \frac{1}{\sqrt{2}}\, \bigl( 0,1,i,0 \bigr) \,,
\nonumber \\
\varepsilon^\alpha_{0} &= \mathcal{N}_\varepsilon\,
   \biggl( q^\alpha - \frac{q^2}{p\cdot q}\, p^\alpha \biggr) \,,
\end{align}
and for the polarization vectors of the $\rho$
\begin{align}
  \label{rho-eps}
e_{+1} &= -\frac{1}{\sqrt{2}}\, 
         \bigl( 0,\cos\Theta, -i, \sin\Theta \bigr) \,,
&
e_{-1} &= \frac{1}{\sqrt{2}}\, 
         \bigl( 0,\cos\Theta, i, \sin\Theta \bigr) \,,
\nonumber \\
e^\alpha_{0} &= \mathcal{N}_e\, \biggl( q'^{\ms\alpha} 
        - \frac{q'^2}{p' \cdot q'}\, p'^{\ms\alpha} \biggr) \,,
\end{align}
where the subscripts indicate helicities.  $\mathcal{N}_\varepsilon$
and $\mathcal{N}_e$ are positive constants ensuring the proper
normalization $\varepsilon_{0}^2 = 1$ and $e_{0}^2 = -1$ of the
longitudinal polarization vectors.  In the $\rho$ rest frame and the
coordinate system $(x,y,z)$ of Fig.~\ref{fig:rho}, our meson
polarization vectors have the standard form $e_{+1} = - (0,1,i,0)
/\sqrt{2}$, $e_{-1} = (0,1,-i,0) /\sqrt{2}$ and $e_{0} = (0,0,0,1)$.
Our phase conventions for the proton and the virtual photon are as in
\cite{Diehl:2005pc}.

We now introduce amplitudes $T^{\ms\nu \sigma}_{\mu \lambda}$ for the
subprocess $\gamma^*(\mu) + p(\lambda) \to \rho(\nu) + p(\sigma)$ with
definite helicities $\mu,\nu,\lambda,\sigma$.  Since the above phase
conventions are defined with reference only to momentum vectors of
this subprocess, the helicity amplitudes only depend on the photon
virtuality, the $\gamma^* p$ scattering energy and the scattering
angle $\Theta$, or equivalently on $Q^2$, $x_B$ and $t = (p-p')^2$.
With our phase conventions they obey the usual parity relations
\begin{equation}
  \label{parity-T}
T^{-\nu -\sigma}_{-\mu -\lambda}
  = (-1)^{\nu -\mu -\sigma + \lambda}\;
    T^{\ms\nu \sigma}_{\mu \lambda}
\end{equation}
for equal $Q^2$, $x_B$ and $t$ on both sides.  With these helicity
amplitudes we define
\begin{equation}
  \label{sdm-def}
\rho^{\ms\nu\nu'}_{\mu\mu', \lambda\lambda'}
 = (N_T + \epsilon N_L)^{-1}\,
   \sum_\sigma T^{\ms\nu \sigma}_{\mu \lambda}\,
               \bigl( T^{\ms\nu'\sigma}_{\mu'\lambda'} \bigr)^* \,.
\end{equation}
Regarding the upper indices this is the spin density matrix of the
vector meson, whereas the lower indices specify the polarizations in
the $\gamma^* p$ state from which the meson is produced.\footnote{%
  Taking the trace in the meson polarization indices we obtain the
  relation $\sum_{\nu} \rho^{\ms\nu\nu}_{\mu\mu', \lambda\lambda'}
  \propto d\sigma_{\mu'\mu}^{\lambda'\lambda}/dt$ between the spin
  density matrix $\rho$ introduced here and the cross sections and
  interference terms used in \protect\cite{Diehl:2005pc}.  Compared
  with \protect\cite{Diehl:2005pc} we take the opposite order of
  indices in $\rho$, so that $\nu$ and $\nu'$ appear in the standard
  order for a spin density matrix.}
The normalization factors
\begin{align}
  \label{norm-def}
N_T &= \half \sum_{\lambda,\nu,\sigma}
       \bigl| T^{\ms\nu\sigma}_{+\lambda} \bigr|^2 \,,
&
N_L &= \half \sum_{\lambda,\nu,\sigma} 
       \bigl| T^{\ms\nu\sigma}_{\0\lambda} \bigr|^2
\end{align}
are proportional to the differential cross sections $d\sigma_T/dt$ and
$d\sigma_L/dt$ for transverse and longitudinal photon polarization,
respectively, and
\begin{equation}
  \label{eps-def}
\epsilon
  = \frac{1 - y - \frac{1}{4}\ms y^2 \gamma^2}{
          1 - y + \frac{1}{2}\ms y^2 + \frac{1}{4}\ms y^2 \gamma^2}
\end{equation}
is the usual ratio of longitudinal and transverse photon flux.  In
addition to $Q^2$, $x_B$ and $t$, the spin density matrix elements
$\rho^{\ms\nu\nu'}_{\mu\mu', \lambda\lambda'}$ depend on $\epsilon$
through the normalization factor $(N_T + \epsilon N_L)$.  If one can
perform a Rosenbluth separation by measuring at different $\epsilon$
but equal $Q^2$ and $x_B$, it is advantageous to normalize them
instead to $N_T$, $N_L$ or $\sqrt{N_T N_L}$ as was done in
\cite{Schilling:1973ag}.  It is straightforward to implement such a
change in the formulae we give in the following.

We find it useful to introduce the combinations
\begin{align}
  \label{sde-1}
\ru{\nu}{\nu'}{\mu}{\mu'} &= \half\bigl( 
   \rho^{\ms\nu\nu'}_{\mu\mu', ++} 
 + \rho^{\ms\nu\nu'}_{\mu\mu', --} \bigr) \,,
& 
\rl{\nu}{\nu'}{\mu}{\mu'} &= \half\bigl( 
   \rho^{\ms\nu\nu'}_{\mu\mu', ++} 
 - \rho^{\ms\nu\nu'}_{\mu\mu', --} \bigr)
\end{align}
for an unpolarized and a longitudinally polarized target, where for
the sake legibility we have labeled the target polarization by $\pm$
instead of $\pm \half$.  The combinations
\begin{align}
  \label{sde-2}
\rs{\nu}{\nu'}{\mu}{\mu'} &= \half\bigl( 
   \rho^{\ms\nu\nu'}_{\mu\mu', +-} 
 + \rho^{\ms\nu\nu'}_{\mu\mu', -+} \bigr) \,,
&
\rn{\nu}{\nu'}{\mu}{\mu'} &= \half\bigl( 
   \rho^{\ms\nu\nu'}_{\mu\mu', +-} 
 - \rho^{\ms\nu\nu'}_{\mu\mu', -+} \bigr)
\end{align}
respectively describe transverse target polarization in the hadron
plane (``sideways'') and perpendicular to it (``normal'').  One
readily finds that the matrices $\ru{}{}{}{}$, $\rl{}{}{}{}$ and
$\rs{}{}{}{}$ are hermitian, whereas $\rn{}{}{}{}$ is antihermitian,
\begin{align}
  \label{rho-hermit}
\ru{\nu'}{\nu}{\mu'}{\mu}
  &= \bigl( \ru{\nu}{\nu'}{\mu}{\mu'} \bigr)^* \,,
&
\rl{\nu'}{\nu}{\mu'}{\mu}
  &= \bigl( \rl{\nu}{\nu'}{\mu}{\mu'} \bigr)^* \,,
&
\rs{\nu'}{\nu}{\mu'}{\mu}
  &= \bigl( \rs{\nu}{\nu'}{\mu}{\mu'} \bigr)^* \,,
\nonumber \\[0.2em]
\rn{\nu'}{\nu}{\mu'}{\mu}
  &= {}-\bigl( \rn{\nu}{\nu'}{\mu}{\mu'} \bigr)^* \,.
\end{align}
The diagonal elements $\ru{\nu}{\nu}{\mu}{\mu}$,
$\rl{\nu}{\nu}{\mu}{\mu}$ and $\rs{\nu}{\nu}{\mu}{\mu}$ are therefore
purely real, whereas $\rn{\nu}{\nu}{\mu}{\mu}$ is purely imaginary.
Furthermore, the parity relations \eqref{parity-T} translate into
\begin{align}
  \label{rho-parity}
\ru{-\nu}{-\nu'}{-\mu}{-\mu'} &=
   (-1)^{\nu-\mu-\nu'+\mu'}\; \ru{\nu}{\nu'}{\mu}{\mu'} \,,
&
\rl{-\nu}{-\nu'}{-\mu}{-\mu'} &=
 - (-1)^{\nu-\mu-\nu'+\mu'}\; \rl{\nu}{\nu'}{\mu}{\mu'} \,,
\nonumber \\
\rn{-\nu}{-\nu'}{-\mu}{-\mu'} &=
   (-1)^{\nu-\mu-\nu'+\mu'}\; \rn{\nu}{\nu'}{\mu}{\mu'} \,,
&
\rs{-\nu}{-\nu'}{-\mu}{-\mu'} &=
 - (-1)^{\nu-\mu-\nu'+\mu'}\; \rs{\nu}{\nu'}{\mu}{\mu'} \,.
\end{align}
As a consequence the matrix elements
\begin{equation}
  \label{real-guys}
\ru{-}{+}{-}{+} \,, \ru{+}{-}{-}{+} \,,
\ru{-}{+}{\0}{\0} \,, \ru{\0}{\0}{-}{+}
\end{equation}
are purely real, whereas the corresponding elements of $\rl{}{}{}{}$,
$\rs{}{}{}{}$ and $\rn{}{}{}{}$ are purely imaginary.

Both experiment and theory indicate that $s$-channel helicity is
approximately conserved in the $\gamma^* \to \rho$ transition for
small invariant momentum transfer $t$.  Correspondingly, one expects
that spin density matrix elements involving the product of two
helicity conserving amplitudes are greater than interference terms
between a helicity conserving and a helicity changing amplitude, and
that those are greater than matrix elements involving the product of
two helicity changing amplitudes (where we refer to the helicities of
the photon and the $\rho$ but not of the nucleon).  Exceptions to this
rule are however possible, since two large amplitudes can have a small
interference term because of their relative phase, and since there can
be cancellation of individually large terms in the linear combinations
\eqref{sde-1} and \eqref{sde-2} associated with different target
polarizations.  With this caveat in mind one can readily assess the
expected size of the spin density matrix elements \eqref{sde-1} and
\eqref{sde-2} by comparing the upper with the lower indices.

Let us now investigate the behavior of our matrix elements for
$\Theta\to 0$, i.e.\ in the limit of forward scattering $\gamma^* p\to
\rho\ms p$.  To this end we perform a partial wave decomposition
\begin{equation}
  \label{pwd}
T^{\nu\sigma}_{\mu\lambda}(\Theta) = \sum_J
t^{\nu\sigma}_{\mu\lambda}(J)\;
  d^J_{\lambda-\mu,\ms \sigma-\nu}(\Theta)
\end{equation}
where we have suppressed the dependence of $T$ and the partial wave
amplitudes $t(J)$ on $Q^2$ and $x_B$.  Using the behavior
$d^J_{m,n}(\Theta) \sim \Theta^{|m-n|}$ of the rotation functions for
$\Theta\to 0$ we readily find
\begin{align}
  \label{theta-lim}
\ru{\nu}{\nu'}{\mu}{\mu'} \,, \rl{\nu}{\nu'}{\mu}{\mu'}
 & \sim \Theta^{\ms p} \,,
&
\rn{\nu}{\nu'}{\mu}{\mu'} \,, \rs{\nu}{\nu'}{\mu}{\mu'}
 & \sim \Theta^{\ms q}
\end{align}
with
\begin{align}
  \label{lim-pow}
p \;\ge\; p_{\text{min}}\;
 &= \min\limits_{\sigma,\ms \lambda \,=\, \pm 1/2}\,
   \Bigl\{\ms \bigl| \nu-\mu-\sigma+\lambda \bigr| 
            + \bigl| \nu'-\mu'-\sigma+\lambda \bigr| \ms\Bigr\} \,,
\nonumber \\
q \;\ge\; q_{\text{min}}\;
 &= \min\limits_{\sigma,\ms \lambda \,=\, \pm 1/2}\,
   \Bigl\{\ms \bigl| \nu-\mu-\sigma+\lambda \bigr| 
            + \bigl| \nu'-\mu'-\sigma-\lambda \bigr| \ms\Bigr\} \,.
\end{align}
With $\Theta \propto (t_0-t)^{1/2}$ for small $\Theta$, we can rewrite
\eqref{theta-lim} as
\begin{align}
  \label{t-lim}
\ru{\nu}{\nu'}{\mu}{\mu'} \,, \rl{\nu}{\nu'}{\mu}{\mu'}
 & \;\underset{t\to t_0}{\sim}\; (t_0-t)^{\ms p/2} \,,
&
\rn{\nu}{\nu'}{\mu}{\mu'} \,, \rs{\nu}{\nu'}{\mu}{\mu'}
 & \;\underset{t\to t_0}{\sim}\; (t_0-t)^{\ms q/2} \,,
\end{align}
where $t_0$ is the value of $t$ for $\Theta=0$ at given $Q^2$ and
$x_B$.  In Tables~\ref{tab:ul} and \ref{tab:ns} we give the
corresponding powers for the linear combinations of spin density
matrix elements that will appear in our results for the cross section
in Section~\ref{sec:cross}.  We have ordered the entries according to
the hierarchy discussed after \eqref{real-guys}, listing first terms
containing the product of two helicity conserving amplitudes, then
terms containing the interference between a helicity conserving and a
helicity changing amplitude, and finally terms which only involve
helicity changing amplitudes (with helicities always referring to the
photon and the $\rho$ but not to the nucleon).

\TABLE{
\caption{\label{tab:ul} Minimum values of the powers which control the
  $t\to t_0$ behavior of combinations of spin density matrix elements
  $\ru{}{}{}{}$ and $\rl{}{}{}{}$ as in \protect\eqref{t-lim}.  Some
  of the combinations are purely real or purely imaginary because of
  the symmetry relations \protect\eqref{rho-hermit} and
  \protect\eqref{rho-parity}, whereas others are complex valued.}
\renewcommand{\arraystretch}{1.2}
\begin{tabular}{llc} \hline\hline
\multicolumn{2}{c}{matrix elements}  &  $p_{\text{min}}$ \\ \hline
$\ru{\0}{\0}{+}{+} + \epsilon \ru{\0}{\0}{\0}{\0}$ \hspace{5.5em} &
\hspace{10.5em} & 0
\\
$\ru{\0}{+}{\0}{+} - \ru{-}{\0}{\0}{+}$ &
$\rl{\0}{+}{\0}{+} - \rl{-}{\0}{\0}{+}$ & 0
\\
$\ru{+}{+}{+}{+} + \ru{-}{-}{+}{+} + 2\epsilon \ru{+}{+}{\0}{\0}$ &
$\rl{+}{+}{+}{+} + \rl{-}{-}{+}{+}$ & 0
\\
$\ru{-}{+}{-}{+}$    & $\rl{-}{+}{-}{+}$    & 0 \\
$\ru{\0}{\0}{\0}{+}$ & $\rl{\0}{\0}{\0}{+}$ & 1 \\
$\ru{\0}{+}{+}{+} - \ru{-}{\0}{+}{+}
    + 2 \re \epsilon \ru{\0}{+}{\0}{\0}$ & 
$\rl{\0}{+}{+}{+} - \rl{-}{\0}{+}{+}
    + 2i \im \epsilon\ms \rl{\0}{+}{\0}{\0}$ & 1 \\
$\ru{\0}{+}{-}{+}$   & $\rl{\0}{+}{-}{+}$ & 1 \\
$\ru{\0}{-}{\0}{+} - \ru{+}{\0}{\0}{+}$ &
$\rl{\0}{-}{\0}{+} - \rl{+}{\0}{\0}{+}$ & 2
\\
$\ru{-}{+}{+}{+} + \epsilon \ru{-}{+}{\0}{\0}$ &
$\rl{-}{+}{+}{+} + \epsilon\ms \rl{-}{+}{\0}{\0}$ & 2
\\
$\ru{+}{+}{-}{+}$ & $\rl{+}{+}{-}{+}$ & 2 \\
$\ru{+}{+}{\0}{+} + \ru{-}{-}{\0}{+}$ &
$\rl{+}{+}{\0}{+} + \rl{-}{-}{\0}{+}$ & 1
\\
$\ru{-}{+}{\0}{+}$  & $\rl{-}{+}{\0}{+}$  & 1 \\
                    & $\rl{\0}{\0}{+}{+}$ & 2 \\ 
$\ru{\0}{\0}{-}{+}$ & $\rl{\0}{\0}{-}{+}$ & 2 \\
$\ru{+}{\0}{-}{+}$  & $\rl{+}{\0}{-}{+}$  & 3 \\
$\ru{+}{-}{\0}{+}$  & $\rl{+}{-}{\0}{+}$  & 3 \\
$\ru{+}{-}{-}{+}$   & $\rl{+}{-}{-}{+}$   & 4 \\ \hline\hline
\end{tabular}
}

\TABLE{
\caption{\label{tab:ns} As Table \protect\ref{tab:ul} but for
  combinations of spin density matrix elements $\rn{}{}{}{}$ and
  $\rs{}{}{}{}$.}
\renewcommand{\arraystretch}{1.2}
\begin{tabular}{llc} \hline\hline
\multicolumn{2}{c}{matrix elements}  &  $q_{\text{min}}$ \\ \hline
$\rn{\0}{\0}{+}{+} + \epsilon \rn{\0}{\0}{\0}{\0}$ \hspace{5.5em} &
\hspace{10.5em} & 1
\\
$\rn{\0}{+}{\0}{+} - \rn{-}{\0}{\0}{+}$ &
$\rs{\0}{+}{\0}{+} - \rs{-}{\0}{\0}{+}$ & 1
\\
$\rn{+}{+}{+}{+} + \rn{-}{-}{+}{+} + 2\epsilon \rn{+}{+}{\0}{\0}$ &
$\rs{+}{+}{+}{+} + \rs{-}{-}{+}{+}$ & 1
\\
$\rn{-}{+}{-}{+}$    & $\rs{-}{+}{-}{+}$    & 1 \\
$\rn{\0}{\0}{\0}{+}$ & $\rs{\0}{\0}{\0}{+}$ & 0 \\
$\rn{\0}{+}{+}{+} - \rn{-}{\0}{+}{+}
    + 2i \im \epsilon \rn{\0}{+}{\0}{\0}$ & 
$\rs{\0}{+}{+}{+} - \rs{-}{\0}{+}{+}
    + 2i \im \epsilon \rs{\0}{+}{\0}{\0}$ & 0 \\
$\rn{\0}{+}{-}{+}$   & $\rs{\0}{+}{-}{+}$ & 0 \\
$\rn{\0}{-}{\0}{+} - \rn{+}{\0}{\0}{+}$ &
$\rs{\0}{-}{\0}{+} - \rs{+}{\0}{\0}{+}$ & 1
\\
$\rn{-}{+}{+}{+} + \epsilon \rn{-}{+}{\0}{\0}$ &
$\rs{-}{+}{+}{+} + \epsilon \rs{-}{+}{\0}{\0}$ & 1
\\
$\rn{+}{+}{-}{+}$ & $\rs{+}{+}{-}{+}$ & 1 \\
$\rn{+}{+}{\0}{+} + \rn{-}{-}{\0}{+}$ &
$\rs{+}{+}{\0}{+} + \rs{-}{-}{\0}{+}$ & 0
\\
$\rn{-}{+}{\0}{+}$  & $\rs{-}{+}{\0}{+}$  & 0 \\
                    & $\rs{\0}{\0}{+}{+}$ & 1 \\
$\rn{\0}{\0}{-}{+}$ & $\rs{\0}{\0}{-}{+}$ & 1 \\
$\rn{+}{\0}{-}{+}$  & $\rs{+}{\0}{-}{+}$  & 2 \\
$\rn{+}{-}{\0}{+}$  & $\rs{+}{-}{\0}{+}$  & 2 \\
$\rn{+}{-}{-}{+}$   & $\rs{+}{-}{-}{+}$   & 3 \\ \hline\hline
\end{tabular}
}

We emphasize that certain partial wave amplitudes
$t^{\nu\sigma}_{\mu\lambda}(J)$ in \eqref{pwd} may be zero or
negligibly small for dynamical reasons.  The actual powers of
$(t_0-t)^{1/2}$ in \eqref{t-lim} can thus be larger than the minimum
values $p_{\text{min}}$ and $q_{\text{min}}$ required by angular
momentum conservation.  If there is for instance no $s$-channel
helicity transferred between the proton-proton and the photon-meson
transitions, then the relevant powers for $\rn{}{}{}{}$ and
$\rs{}{}{}{}$ are given by $q = p_{\text{min}} + 1$, which is equal to
$q_{\text{min}} + 2$ for all but the first four entries in
Tables~\ref{tab:ul} and~\ref{tab:ns}.  A concrete realization of this
scenario is the calculation in \cite{Goloskokov:2005sd}, where the
proton-proton transition is described by the generalized parton
distributions $H$, $E$ and $\tilde{H}$, $\tilde{E}$, which do not
allow for helicity transfer to the photon-meson transition.

In the limit of large $Q^2$ at fixed $x_B$ and $t$, the proof of the
factorization theorem in \cite{Collins:1996fb} implies that the
transition from a longitudinal photon to a longitudinal $\rho$ becomes
dominant, with all other transitions suppressed by powers of $1/Q$.
In this limit only the spin density matrix elements
$\ru{\0}{\0}{\0}{\0}$ and $\rn{\0}{\0}{\0}{\0}$ survive and can be
expressed as convolutions of hard-scattering kernels with generalized
parton distributions and the light-cone distribution amplitude of the
$\rho$.  To leading order in $1/Q$ one has in particular
\begin{equation}
\frac{\im\rn{\0}{\0}{\0}{\0}}{\ru{\0}{\0}{\0}{\0}}
 = \frac{\sqrt{t_0-t\rule{0pt}{1.8ex}}}{\rule{0pt}{0.85em}M_N}\,
   \frac{\sqrt{1-\xi^2}\,
     \im\bigl( \mathcal{E}^* \mathcal{H} \bigr)}{\rule{0pt}{1em}
   (1-\xi^2)\, | \mathcal{H} |^2
   - \bigl( \xi^2 + t/(4M_N^2) \bigr)\, | \mathcal{E} |^2
   - 2\xi^2 \re \bigl( \mathcal{E}^* \mathcal{H} \bigr)} \,,
\end{equation}
where $\xi = x_B /(2-x_B)$ and the convolution integrals $\mathcal{H}$
and $\mathcal{E}$ are for instance given in \cite{Diehl:2005pc}.
Experimental results and phenomenological analysis show however that
$1/Q^2$ suppressed effects can be numerically significant for $Q^2$ of
several $\gev^2$, see e.g.\ \cite{Vanderhaeghen:1999xj,%
Goloskokov:2005sd,Diehl:2005gn,Goloskokov:2006hr}.  This concerns both
power corrections within $\ru{\0}{\0}{\0}{\0}$ or
$\rn{\0}{\0}{\0}{\0}$ and formally power suppressed spin density
matrix elements such as $\ru{+}{+}{+}{+}$ or $\ru{\0}{+}{\0}{+}$.
The detailed analysis in \cite{Collins:1996fb} reveals that beyond
leading-power accuracy in $1/Q$, factorization of meson production
into a hard-scattering subprocess and nonperturbative quantities
pertaining either to the target or to the meson may be broken.  On the
other hand, factorization based approaches which go beyond leading
power in $1/Q$ and in particular also evaluate transition amplitudes
for transverse polarization of the $\gamma^*$ or $\rho$ have been
phenomenologically rather successful, see e.g.\
\cite{Ivanov:1998gk,Goloskokov:2005sd}

Let us finally generalize our considerations to the process
\begin{equation}
e(l) + p(p) \to e(l') + X(p') + \rho(q') \,,
\end{equation}
where the target proton dissociates into a hadronic system $X$.  In
analogy to the elastic case one can introduce helicity amplitudes
$T^{\nu\sigma,\ms X}_{\mu\lambda}$ and combine them into spin density
matrix elements
\begin{equation}
\rho^{\ms\nu\nu'}_{\mu\mu', \lambda\lambda'}
 = (N_T + \epsilon N_L)^{-1}\,
 \sum_{X,\ms \sigma} T^{\ms\nu \sigma,\ms X}_{\mu \lambda}\,
              \bigl( T^{\ms\nu'\sigma,\ms X}_{\mu'\lambda'} \bigr)^* \,.
\end{equation}
The normalization factors $N_T$ and $N_L$ are defined as in
\eqref{norm-def} but with an additional sum over all hadronic states
$X$ of given invariant mass $M_X$, on which
$\rho^{\ms\nu\nu'}_{\mu\mu', \lambda\lambda'}$ now depends in addition
to $Q^2$, $x_B$, $t$ and $\epsilon$.  The combinations \eqref{sde-1}
and \eqref{sde-2} for different target polarization have the same
symmetry properties \eqref{rho-hermit} and \eqref{rho-parity} as in
the elastic case.  Their behavior for $t\to t_0$ can be different,
since in \eqref{lim-pow} one must now take the minimum over all
possible helicities $\sigma= \pm \half, \pm \frac{3}{2}, \ldots$ of
the hadronic system $X$.  One finds however that the powers
$p_{\text{min}}$ and $q_{\text{min}}$ for the combinations of spin
density matrix elements in Tables~\ref{tab:ul} and \ref{tab:ns} are
the same as in the elastic case.  The results in the remainder of this
work only depend on the properties \eqref{rho-hermit} and
\eqref{rho-parity} and thus immediately generalize to the case of
target dissociation.


\section{The angular distribution}
\label{sec:cross}

The calculation of the cross section for $ep \to ep\, \pi^+\pi^-$
proceeds by using standard methods and we shall only sketch the
essential steps.  More details are for instance given
in~\cite{Schilling:1973ag,Arens:1996xw,Diehl:2005pc}.  With our phase
conventions the polarization state of the proton target is described
by the spin density matrix
\begin{equation}
\tau_{\lambda\lambda'} = \frac{1}{2} 
\begin{pmatrix}
  1 + S_L & S_T\ms e^{-i(\phi-\phi_S)} \\[0.1em]
  S_T\ms e^{i(\phi-\phi_S)} & 1 - S_L
\end{pmatrix} ,
\end{equation}
which is to be contracted with the matrix in \eqref{sdm-def}.  The
result is conveniently expressed in terms of the combinations
\eqref{sde-1} and \eqref{sde-2} as
\begin{align}
  \label{contract-1}
\sum_{\lambda,\lambda'} \tau^{}_{\lambda\lambda'}\, 
  \rho^{\nu\nu'}_{\mu\mu', \lambda\lambda'}
 &= \ru{\nu}{\nu'}{\mu}{\mu'} 
    + S_L^{}\ms \rl{\nu}{\nu'}{\mu}{\mu'}
    + S_T^{} \cos(\phi-\phi_S)\, \rs{\nu}{\nu'}{\mu}{\mu'}
    - S_T^{} \sin(\phi-\phi_S)\, i \rn{\nu}{\nu'}{\mu}{\mu'}
\end{align}
and describes the subprocess $\gamma^* p\to \rho\ms p$.  The decay
$\rho\to \pi^+\pi^-$ is taken into account by multiplication with the
spherical harmonics,
\begin{equation}
  \label{contract-2}
\rho^{}_{\mu\mu'} =
  \sum_{\nu,\nu'} 
  \sum_{\lambda,\lambda'} \tau^{}_{\lambda\lambda'}\, 
  \rho^{\nu\nu'}_{\mu\mu', \lambda\lambda'}\,
  Y^{}_{1\ms \nu}(\varphi,\vartheta)\,
  Y^*_{1\ms \nu'}(\varphi,\vartheta) \,,
\end{equation}
where
\begin{align}
Y_{1 +1} &= - \sqrt{\frac{3}{8\pi}}\, \sin\vartheta\; e^{i\varphi} \,,
&
Y_{1 0} &= \sqrt{\frac{3}{4\pi}}\, \cos\vartheta \,,
&
Y_{1 -1} &= \sqrt{\frac{3}{8\pi}}\, \sin\vartheta\; e^{-i\varphi} \,.
\end{align}
To obtain the cross section for the overall process $ep\to ep\ms
\pi^+\pi^-$ one must finally contract the matrix $\rho_{\mu\mu'}$ in
\eqref{contract-2} with the spin density matrix of the virtual
photon.\footnote{%
Up to a global factor, the result of this contraction can e.g.\ be
obtained from eq.~(3.20) of \protect\cite{Arens:1996xw}, with
$\rho_{\mu\mu'}$ in the present work corresponding to
$\sigma_{\mu'\mu}^{\smash{(X)}}$ in \protect\cite{Arens:1996xw} and
$\phi_{\ms\text{here}}^{} =
-\varphi_{\ms\text{\protect\cite{Arens:1996xw}}}$.}
The cross section can be written as
\begin{align}
  \label{X-sect}
& \frac{d\sigma}{d\psi\, d\phi\, d\varphi\, d(\cos\vartheta)\,
                 dx_B\, dQ^2\, dt}
 = \frac{1}{(2\pi)^2}\, \frac{d\sigma}{dx_B\, dQ^2\, dt}\,
\nonumber \\[0.3em]
&\qquad\qquad \times
   \Bigl( W_{UU} + P_\ell\ms W_{LU} 
        + S_L W_{UL} + P_\ell\ms S_L W_{LL}
        + S_T W_{UT} + P_\ell\ms S_T W_{LT} \Bigr)
\end{align}
with
\begin{equation}
\frac{d\sigma}{dx_B\, dQ^2\, dt} =
\frac{\alpha_{\rm em}}{2\pi}\, \frac{y^2}{1-\epsilon}\,
   \frac{1-x_B}{x_B}\, \frac{1}{Q^2}\,
   \biggl( \frac{d\sigma_T}{dt} 
         + \epsilon\, \frac{d\sigma_L}{dt} \biggr) \,,
\end{equation}
where $d\sigma_T/dt$ and $d\sigma_L/dt$ are the usual $\gamma^* p$
cross sections for a transverse and longitudinal photon and an
unpolarized proton, with Hand's convention for virtual photon flux.
The angular distribution is described by the quantities $W_{XY}$,
where $X$ specifies the beam and $Y$ the target polarization.  The
normalization of the unpolarized term $W_{UU}$ is
\begin{equation}
  \label{W-norm}
\int \frac{d\phi}{2\pi}
\int d\varphi\; d(\cos\vartheta)\; W_{UU}(\phi,\varphi,\vartheta)
  = 1 \,.
\end{equation}
To limit the length of subsequent expressions, we further decompose
the coefficients according to the $\rho$ polarization and write
\begin{align}
  \label{W-dec}
& W_{XY}(\phi,\varphi,\vartheta) 
\nonumber \\
& = \frac{3}{4\pi} \biggl[\,
     \cos^2\bs\vartheta\; W_{XY}^{LL}(\phi)
   + \sqrt{2} \cos\vartheta\, \sin\vartheta\;
                          W_{XY}^{LT}(\phi,\varphi)
   + \sin^2\bs\vartheta\; W_{XY}^{TT}(\phi,\varphi)
   \,\biggr]
\end{align}
for $X,Y = U,L$.  The production of a longitudinal $\rho$ is described
\pagebreak[2]
by $W_{XY}^{LL}$, the production of a transverse $\rho$ (including the
interference between positive and negative $\rho$ helicity) by
$W_{XY}^{TT}$, and the interference between longitudinal and
transverse $\rho$ polarization by $W_{XY}^{LT}$.  For a transversely
polarized target we have in addition a dependence on $\phi_S$,
\begin{align}
& W_{XT}(\phi_S,\phi,\varphi,\vartheta) 
\nonumber \\
& = \frac{3}{4\pi} \biggl[\,
     \cos^2\bs\vartheta\; W_{XT}^{LL}(\phi_S,\phi)
   + \sqrt{2} \cos\vartheta\, \sin\vartheta\;
                          W_{XT}^{LT}(\phi_S,\phi,\varphi)
   + \sin^2\bs\vartheta\; W_{XT}^{TT}(\phi_S,\phi,\varphi)
   \,\biggr]
\end{align}
with $X = U,L$.  In addition to the angles, all coefficients $W_{XY}$
depend on $Q^2$, $x_B$ and $t$, which we have not displayed for the
sake of legibility.

For unpolarized target and beam we have
%
\begin{align}
  \label{WUU}
W_{UU}^{LL}(\phi) &=
  \bigl( \ru{\0}{\0}{+}{+} + \epsilon \ru{\0}{\0}{\0}{\0} \bigr)
  - 2\cos\phi\, \sqrt{\epsilon (1+\epsilon)}\ms \re\ru{\0}{\0}{\0}{+}
  - \cos(2\phi)\, \epsilon \ru{\0}{\0}{-}{+} \,,
  \phantom{\frac{1}{2}}
\nonumber \\
W_{UU}^{LT}(\phi,\varphi) &=
  \cos(\phi + \varphi)\, \sqrt{\epsilon (1+\epsilon)}\ms
  \re\bigl( \ru{\0}{+}{\0}{+} - \ru{-}{\0}{\0}{+} \bigr)
  \phantom{\frac{1}{2}}
\nonumber \\
& \quad{}
  - \cos\varphi\,
  \re\bigl( \ru{\0}{+}{+}{+} - \ru{-}{\0}{+}{+}
          + 2\epsilon \ru{\0}{+}{\0}{\0} \bigr)
  + \cos(2\phi + \varphi)\, \epsilon \re\ru{\0}{+}{-}{+}
  \phantom{\frac{1}{2}}
\nonumber \\
& \quad{}
  - \cos(\phi - \varphi)\, \sqrt{\epsilon (1+\epsilon)}\ms
    \re\bigl( \ru{\0}{-}{\0}{+} - \ru{+}{\0}{\0}{+} \bigr)
  + \cos(2\phi - \varphi)\, \epsilon \re\ru{+}{\0}{-}{+} \,,
  \phantom{\frac{1}{2}}
\nonumber \\
W_{UU}^{TT}(\phi,\varphi) &=
  \tfrac{1}{2}\ms \bigl( \ru{+}{+}{+}{+} + \ru{-}{-}{+}{+}
                      + 2\epsilon \ru{+}{+}{\0}{\0} \bigr)
  + \tfrac{1}{2} \cos(2\phi + 2\varphi)\, \epsilon \ru{-}{+}{-}{+}
  \phantom{\frac{1}{2}}
\nonumber \\
& \quad{}
  - \cos\phi\, \sqrt{\epsilon (1+\epsilon)}\ms
    \re\bigl( \ru{+}{+}{\0}{+} + \ru{-}{-}{\0}{+} \bigr)
  + \cos(\phi + 2\varphi)\, \sqrt{\epsilon (1+\epsilon)}\ms
    \re\ru{-}{+}{\0}{+}
  \phantom{\frac{1}{2}}
\nonumber \\
& \quad{}
  - \cos(2\varphi)\,
    \re\bigl( \ru{-}{+}{+}{+} + \epsilon \ru{-}{+}{\0}{\0} \bigr)
  - \cos(2\phi)\, \epsilon \re\ru{+}{+}{-}{+}
  \phantom{\frac{1}{2}}
\nonumber \\
& \quad{}
  + \cos(\phi - 2\varphi)\, \sqrt{\epsilon (1+\epsilon)}\ms
    \re\ru{+}{-}{\0}{+}
  + \tfrac{1}{2} \cos(2\phi - 2\varphi)\, \epsilon \ru{+}{-}{-}{+} \,.
  \phantom{\frac{1}{2}}
\end{align}
Here and in the following we order terms according to the hierarchy
discussed after \eqref{real-guys}, as already done in
Table~\ref{tab:ul}.
The terms independent of $\phi$ and $\varphi$ in $W_{UU}^{LL}$ and
$W_{UU}^{TT}$ are related by
\begin{equation}
  \label{norm-rel}
\ru{+}{+}{+}{+} + \ru{-}{-}{+}{+} + 2\epsilon \ru{+}{+}{\0}{\0}
= 1 - 
  \bigl( \ru{\0}{\0}{+}{+} + \epsilon \ru{\0}{\0}{\0}{\0} \bigr) \,,
\end{equation}
which ensures the normalization condition \eqref{W-norm}.  The terms
for beam polarization with an unpolarized target read
%
\begin{align}
  \label{WLU}
W_{LU}^{LL}(\phi) &=
  - 2 \sin\phi\, \sqrt{\epsilon (1-\epsilon)}\ms \im\ru{\0}{\0}{\0}{+}
  \,,
  \phantom{\frac{1}{2}}
\nonumber \\
W_{LU}^{LT}(\phi,\varphi) &=
  \sin(\phi + \varphi)\, \sqrt{\epsilon (1-\epsilon)}\ms
    \im\bigl( \ru{\0}{+}{\0}{+} - \ru{-}{\0}{\0}{+} \bigr)
  \phantom{\frac{1}{2}}
\nonumber \\
& \quad{}
  - \sin\varphi\, \sqrt{1 - \epsilon^2}\,
    \im\bigl( \ru{\0}{+}{+}{+} - \ru{-}{\0}{+}{+} \bigr)
  \phantom{\frac{1}{2}}
\nonumber \\
& \quad{}
  - \sin(\phi - \varphi)\, \sqrt{\epsilon (1-\epsilon)}\ms
    \im\bigl( \ru{\0}{-}{\0}{+} - \ru{+}{\0}{\0}{+} \bigr) \,,
  \phantom{\frac{1}{2}}
\nonumber \\
W_{LU}^{TT}(\phi,\varphi) &=
  - \sin\phi\, \sqrt{\epsilon (1-\epsilon)}\ms
    \im\bigl( \ru{+}{+}{\0}{+} + \ru{-}{-}{\0}{+} \bigr)
  + \sin(\phi + 2\varphi)\, \sqrt{\epsilon (1-\epsilon)}\ms
    \im\ru{-}{+}{\0}{+}
  \phantom{\frac{1}{2}}
\nonumber \\
& \quad{}
  - \sin(2\varphi)\, \sqrt{1 - \epsilon^2}\, \im\ru{-}{+}{+}{+}
  \phantom{\frac{1}{2}}
\nonumber \\
& \quad{}
  + \sin(\phi - 2\varphi)\, \sqrt{\epsilon (1-\epsilon)}\ms
    \im\ru{+}{-}{\0}{+} \,.
  \phantom{\frac{1}{2}}
\end{align}
The results for longitudinal target polarization are very similar,
with
\begin{align}
  \label{WUL}
W_{UL}^{LL}(\phi) &=
  - 2 \sin\phi\, \sqrt{\epsilon (1+\epsilon)}\ms \im\rl{\0}{\0}{\0}{+}
  - \sin(2\phi)\, \epsilon \im\rl{\0}{\0}{-}{+} \,,
  \phantom{\frac{1}{2}}
\nonumber \\
W_{UL}^{LT}(\phi,\varphi) &=
  \sin(\phi + \varphi)\, \sqrt{\epsilon (1+\epsilon)}\ms
  \im\bigl( \rl{\0}{+}{\0}{+} - \rl{-}{\0}{\0}{+} \bigr)
  \phantom{\frac{1}{2}}
\nonumber \\
& \quad{}
  - \sin\varphi\,
  \im\bigl( \rl{\0}{+}{+}{+} - \rl{-}{\0}{+}{+}
          + 2\epsilon\ms \rl{\0}{+}{\0}{\0} \bigr)
  + \sin(2\phi + \varphi)\, \epsilon \im\rl{\0}{+}{-}{+}
  \phantom{\frac{1}{2}}
\nonumber \\
& \quad{}
  - \sin(\phi - \varphi)\, \sqrt{\epsilon (1+\epsilon)}\ms
    \im\bigl( \rl{\0}{-}{\0}{+} - \rl{+}{\0}{\0}{+} \bigr)
  + \sin(2\phi - \varphi)\, \epsilon \im\rl{+}{\0}{-}{+} \,,
  \phantom{\frac{1}{2}}
\nonumber \\
W_{UL}^{TT}(\phi,\varphi) &=
  \tfrac{1}{2} \sin(2\phi + 2\varphi)\, \epsilon \im\rl{-}{+}{-}{+}
  \phantom{\frac{1}{2}}
\nonumber \\
& \quad{}
  - \sin\phi\, \sqrt{\epsilon (1+\epsilon)}\ms
    \im\bigl( \rl{+}{+}{\0}{+} + \rl{-}{-}{\0}{+} \bigr)
  + \sin(\phi + 2\varphi)\, \sqrt{\epsilon (1+\epsilon)}\ms
    \im\rl{-}{+}{\0}{+}
  \phantom{\frac{1}{2}}
\nonumber \\
& \quad{}
  - \sin(2\varphi)\,
    \im\bigl( \rl{-}{+}{+}{+} + \epsilon\ms \rl{-}{+}{\0}{\0} \bigr)
  - \sin(2\phi)\, \epsilon \im\rl{+}{+}{-}{+}
  \phantom{\frac{1}{2}}
\nonumber \\
& \quad{}
  + \sin(\phi - 2\varphi)\, \sqrt{\epsilon (1+\epsilon)}\ms
    \im\rl{+}{-}{\0}{+}
  + \tfrac{1}{2} \sin(2\phi - 2\varphi)\, \epsilon \im\rl{+}{-}{-}{+}
  \phantom{\frac{1}{2}}
\end{align}
for an unpolarized beam, and
%
\begin{align}
  \label{WLL}
W_{LL}^{LL}(\phi) &=
  - 2\cos\phi\, \sqrt{\epsilon (1-\epsilon)}\ms \re\rl{\0}{\0}{\0}{+}
  + \sqrt{1 - \epsilon^2}\; \rl{\0}{\0}{+}{+} \,,
  \phantom{\frac{1}{2}}
\nonumber \\
W_{LL}^{LT}(\phi,\varphi) &=
  \cos(\phi + \varphi)\, \sqrt{\epsilon (1-\epsilon)}\ms
  \re\bigl( \rl{\0}{+}{\0}{+} - \rl{-}{\0}{\0}{+} \bigr)
  \phantom{\frac{1}{2}}
\nonumber \\
& \quad{}
  - \cos\varphi\, \sqrt{1 - \epsilon^2}\,
    \re\bigl( \rl{\0}{+}{+}{+} - \rl{-}{\0}{+}{+} \bigr)
  \phantom{\frac{1}{2}}
\nonumber \\
& \quad{}
  - \cos(\phi - \varphi)\, \sqrt{\epsilon (1-\epsilon)}\ms
    \re\bigl( \rl{\0}{-}{\0}{+} - \rl{+}{\0}{\0}{+} \bigr) \,,
  \phantom{\frac{1}{2}}
\nonumber \\
W_{LL}^{TT}(\phi,\varphi) &=
  \sqrt{1 - \epsilon^2}\; \tfrac{1}{2}\ms
  \bigl( \rl{+}{+}{+}{+} + \rl{-}{-}{+}{+} \bigr)
  \phantom{\frac{1}{2}}
\nonumber \\
& \quad{}
  - \cos\phi\, \sqrt{\epsilon (1-\epsilon)}\ms
    \re\bigl( \rl{+}{+}{\0}{+} + \rl{-}{-}{\0}{+} \bigr)
  + \cos(\phi + 2\varphi)\, \sqrt{\epsilon (1-\epsilon)}\ms
    \re\rl{-}{+}{\0}{+}
  \phantom{\frac{1}{2}}
\nonumber \\
& \quad{}
  - \cos(2\varphi)\, \sqrt{1 - \epsilon^2}\, \re\rl{-}{+}{+}{+}
\nonumber \\
& \quad{}
  + \cos(\phi - 2\varphi)\, \sqrt{\epsilon (1-\epsilon)}\ms
    \re\rl{+}{-}{\0}{+}
  \phantom{\frac{1}{2}}
\end{align}
for beam polarization.  In \eqref{WUU} to \eqref{WLL} we have used the
symmetry relations \eqref{rho-hermit} and \eqref{rho-parity} to write
our results with a minimal set of matrix elements
$\ru{\nu}{\nu'}{\mu}{\mu'}$ or $\rl{\nu}{\nu'}{\mu}{\mu'}$.  Although
they are a little lengthy, their structure is quite simple:
\begin{enumerate}
\item The combinations $\ru{+}{+}{\mu}{\mu'} + \ru{-}{-}{\mu}{\mu'}$,
  $\ru{\0}{+}{\mu}{\mu'} - \ru{-}{\0}{\mu}{\mu'}$ and
  $\ru{\0}{-}{\mu}{\mu'} - \ru{+}{\0}{\mu}{\mu'}$ and their analogs
  for $\rl{}{}{}{}$ always appear together because the corresponding
  products of spherical harmonics are identical, $Y_{1 +1}^{} Y_{1
  +1}^* = Y_{1 -1}^{} Y_{1 -1}^*$ and $Y_{1 0}^{} Y_{1 +1}^* = - Y_{1
  -1}^{} Y_{1 0}^*$.  In some cases the corresponding sum can be
  simplified using symmetry relations like $\ru{+}{+}{\0}{\0} +
  \ru{-}{-}{\0}{\0} = 2 \ru{+}{+}{\0}{\0}$, but in others one remains
  with a linear combination of matrix elements that cannot be
  separated.  With the caveats discussed after \eqref{real-guys} one
  finds however that these combinations are dominated by a single
  term.  Exceptions are $\re\bigl( \ru{\0}{+}{+}{+} - \ru{-}{\0}{+}{+}
  + 2\epsilon \ru{\0}{+}{\0}{\0} \bigr)$ and $\im\bigl(
  \rl{\0}{+}{+}{+} - \rl{-}{\0}{+}{+} + 2\epsilon\ms
  \rl{\0}{+}{\0}{\0} \bigr)$, each of which contains two interference
  terms between a helicity conserving and a helicity changing
  amplitude.
\item An angular dependence through $(k\phi + m\varphi)$ is associated
  with the interference between transverse and longitudinal $\rho$
  polarization for $|m|=1$, the interference between positive and
  negative $\rho$ helicity for $|m|=2$, and equal $\rho$ polarization
  in the amplitude and its conjugate for $m=0$.  In the same way
  $|k|=1$, $|k|=2$ and $k=0$ are related to the virtual photon
  polarization.  Notice that for $m=0$ one can distinguish transverse
  and longitudinal $\rho$ production by the $\vartheta$ dependence in
  \eqref{W-dec}, whereas for $k=0$ the separation of terms for
  transverse and longitudinal photons requires variation of
  $\epsilon$.  The beam spin asymmetries $W_{LU}$ and $W_{LL}$ contain
  no terms with $|k|=2$, because there is no term with $P_\ell \cos
  2\phi$ or $P_\ell \sin 2\phi$ in the spin density matrix of the
  virtual photon.
\item The unpolarized or doubly polarized terms $W_{UU}$ and $W_{LL}$
  depend on $\re\ru{}{}{}{}$ or $\re\rl{}{}{}{}$ and are even under
  the reflection $(\phi,\varphi) \to (-\phi,-\varphi)$ of the
  azimuthal angles, whereas the single spin asymmetries $W_{LU}$ and
  $W_{UL}$ depend on $\im\ru{}{}{}{}$ or $\im\rl{}{}{}{}$ and are odd
  under $(\phi,\varphi) \to (-\phi,-\varphi)$.  This is a consequence
  of parity and time reversal invariance.
\item As we have written our results, the angular distribution for
  longitudinal target polarization can be obtained from the one for an
  unpolarized target by replacing
\begin{align}
  \label{repl-rules-lu}
\cos(k\phi + m\varphi)\, \re\ru{}{}{}{} &\to 
\sin(k\phi + m\varphi)\, \im\rl{}{}{}{} \,,
\nonumber \\
\sin(k\phi + m\varphi)\, \im\ru{}{}{}{} &\to
\cos(k\phi + m\varphi)\, \re\rl{}{}{}{} \,.
\end{align}
  Terms with $k=m=0$ in $W_{UU}$ and $W_{LL}$ are independent of
  $\phi$ and $\varphi$, and have of course no counterparts in $W_{UL}$
  or $W_{LU}$.  This corresponds to 16 terms with a different angular
  dependence in $W_{UU}$ and 14 terms in $W_{UL}$, and to 10 terms in
  $W_{LL}$ and 8 terms in $W_{LU}$.
\end{enumerate}

The symmetry properties \eqref{rho-hermit} and \eqref{rho-parity},
which we used to obtain \eqref{WUU} to \eqref{WLL}, are identical for
$\ru{\nu}{\nu'}{\mu}{\mu'}$ and $i \rn{\nu}{\nu'}{\mu}{\mu'}$, as well
as for $\rl{\nu}{\nu'}{\mu}{\mu'}$ and $\rs{\nu}{\nu'}{\mu}{\mu'}$.
According to \eqref{contract-1} the cross section for a transversely
polarized target can therefore be obtained from the one for
longitudinal and no target polarization by the replacements
\begin{align}
  \label{repl-rules}
\re \ru{}{}{}{} &\to \phantom{-}S_T 
                     \sin(\phi-\phi_S)\, \im \rn{}{}{}{} \,,
&
S_L \im \rl{}{}{}{} &\to S_T \cos(\phi-\phi_S)\, \im \rs{}{}{}{} \,,
\nonumber \\
\im \ru{}{}{}{} &\to -S_T \sin(\phi-\phi_S)\, \re \rn{}{}{}{} \,,
&
S_L \re \rl{}{}{}{} &\to S_T \cos(\phi-\phi_S)\, \re \rs{}{}{}{} \,.
\end{align}
We thus simply have
%
\begin{align}
  \label{WUT}
W_{UT}^{LL}(\phi_S,\phi) &=
\sin(\phi-\phi_S)\, \Bigl[\,
  \im\bigl( \rn{\0}{\0}{+}{+} + \epsilon \rn{\0}{\0}{\0}{\0} \bigr)
  \phantom{\frac{1}{2}}
\nonumber \\
& \qquad{}
  - 2\cos\phi\, \sqrt{\epsilon (1+\epsilon)}\ms \im\rn{\0}{\0}{\0}{+}
  - \cos(2\phi)\, \epsilon \im\rn{\0}{\0}{-}{+}
\,\Bigr]
  \phantom{\frac{1}{2}}
\nonumber \\
& {}
+ \cos(\phi-\phi_S)\, \Bigl[\,
  - 2\sin\phi\, \sqrt{\epsilon (1+\epsilon)}\ms \im\rs{\0}{\0}{\0}{+}
  - \sin(2\phi)\, \epsilon \im\rs{\0}{\0}{-}{+}
\,\Bigr] \,,
  \phantom{\frac{1}{2}}
\nonumber \\
W_{UT}^{LT}(\phi_S,\phi,\varphi) &=
\sin(\phi-\phi_S)\, \Bigl[\,
  \cos(\phi + \varphi)\, \sqrt{\epsilon (1+\epsilon)}\ms
  \im\bigl( \rn{\0}{+}{\0}{+} - \rn{-}{\0}{\0}{+} \bigr)
  \phantom{\frac{1}{2}}
\nonumber \\
& \qquad{}
  - \cos\varphi\,
  \im\bigl( \rn{\0}{+}{+}{+} - \rn{-}{\0}{+}{+}
          + 2\epsilon \rn{\0}{+}{\0}{\0} \bigr)
  + \cos(2\phi + \varphi)\, \epsilon \im\rn{\0}{+}{-}{+}
  \phantom{\frac{1}{2}}
\nonumber \\
& \qquad{}
  - \cos(\phi - \varphi)\, \sqrt{\epsilon (1+\epsilon)}\ms
    \im\bigl( \rn{\0}{-}{\0}{+} - \rn{+}{\0}{\0}{+} \bigr)
  + \cos(2\phi - \varphi)\, \epsilon \im\rn{+}{\0}{-}{+}
\,\Bigr]
  \phantom{\frac{1}{2}}
\displaybreak[2]
\nonumber \\
& {}
+ \cos(\phi-\phi_S)\, \Bigl[\,
  \sin(\phi + \varphi)\, \sqrt{\epsilon (1+\epsilon)}\ms
  \im\bigl( \rs{\0}{+}{\0}{+} - \rs{-}{\0}{\0}{+} \bigr)
  \phantom{\frac{1}{2}}
\nonumber \\
& \qquad{}
  - \sin\varphi\,
  \im\bigl( \rs{\0}{+}{+}{+} - \rs{-}{\0}{+}{+}
          + 2\epsilon \rs{\0}{+}{\0}{\0} \bigr)
  + \sin(2\phi + \varphi)\, \epsilon \im\rs{\0}{+}{-}{+}
  \phantom{\frac{1}{2}}
\nonumber \\
& \qquad{}
  - \sin(\phi - \varphi)\, \sqrt{\epsilon (1+\epsilon)}\ms
    \im\bigl( \rs{\0}{-}{\0}{+} - \rs{+}{\0}{\0}{+} \bigr)
  + \sin(2\phi - \varphi)\, \epsilon \im\rs{+}{\0}{-}{+}
\,\Bigr] \,,
  \phantom{\frac{1}{2}}
\nonumber \\
W_{UT}^{TT}(\phi_S,\phi,\varphi) &=
\sin(\phi-\phi_S)\, \Bigl[\,
  \tfrac{1}{2} \im\bigl( \rn{+}{+}{+}{+} + \rn{-}{-}{+}{+}
                      + 2\epsilon \rn{+}{+}{\0}{\0} \bigr)
  + \tfrac{1}{2} \cos(2\phi + 2\varphi)\, \epsilon \im\rn{-}{+}{-}{+}
  \phantom{\frac{1}{2}}
\nonumber \\
& \qquad{}
  - \cos\phi\, \sqrt{\epsilon (1+\epsilon)}\ms
    \im\bigl( \rn{+}{+}{\0}{+} + \rn{-}{-}{\0}{+} \bigr)
  + \cos(\phi + 2\varphi)\, \sqrt{\epsilon (1+\epsilon)}\ms
    \im\rn{-}{+}{\0}{+}
  \phantom{\frac{1}{2}}
\nonumber \\
& \qquad{}
  - \cos(2\varphi)\,
    \im\bigl( \rn{-}{+}{+}{+} + \epsilon \rn{-}{+}{\0}{\0} \bigr)
  - \cos(2\phi)\, \epsilon \im\rn{+}{+}{-}{+}
  \phantom{\frac{1}{2}}
\nonumber \\
& \qquad{}
  + \cos(\phi - 2\varphi)\, \sqrt{\epsilon (1+\epsilon)}\ms
    \im\rn{+}{-}{\0}{+}
  + \tfrac{1}{2} \cos(2\phi - 2\varphi)\, \epsilon \im\rn{+}{-}{-}{+}
\,\Bigr]
  \phantom{\frac{1}{2}}
\nonumber \\
& {}
+ \cos(\phi-\phi_S)\, \Bigl[\,
  \tfrac{1}{2} \sin(2\phi + 2\varphi)\, \epsilon \im\rs{-}{+}{-}{+}
  \phantom{\frac{1}{2}}
\nonumber \\
& \qquad{}
  - \sin\phi\,  \sqrt{\epsilon (1+\epsilon)}\ms
    \im\bigl( \rs{+}{+}{\0}{+} + \rs{-}{-}{\0}{+} \bigr)
  + \sin(\phi + 2\varphi)\, \sqrt{\epsilon (1+\epsilon)}\ms
    \im\rs{-}{+}{\0}{+}
  \phantom{\frac{1}{2}}
\nonumber \\
& \qquad{}
  - \sin(2\varphi)\,
    \im\bigl( \rs{-}{+}{+}{+} + \epsilon \rs{-}{+}{\0}{\0} \bigr)
  - \sin(2\phi)\, \epsilon \im\rs{+}{+}{-}{+}
  \phantom{\frac{1}{2}}
\nonumber \\
& \qquad{}
  + \sin(\phi - 2\varphi)\, \sqrt{\epsilon (1+\epsilon)}\ms
    \im\rs{+}{-}{\0}{+}
  + \tfrac{1}{2} \sin(2\phi - 2\varphi)\, \epsilon \im\rs{+}{-}{-}{+}
\,\Bigr]
  \phantom{\frac{1}{2}}
\end{align}
for an unpolarized beam, and
%
\begin{align}
  \label{WLT}
W_{LT}^{LL}(\phi_S,\phi) &=
\sin(\phi-\phi_S)\, \Bigl[\,\,
  2\sin\phi\, \sqrt{\epsilon (1-\epsilon)}\ms \re\rn{\0}{\0}{\0}{+}
\,\Bigr]
  \phantom{\frac{1}{2}}
\nonumber \\
& {}
+ \cos(\phi-\phi_S)\, \Bigl[\,
  - 2\cos\phi\, \sqrt{\epsilon (1-\epsilon)}\ms \re\rs{\0}{\0}{\0}{+}
  + \sqrt{1 - \epsilon^2}\, \rs{\0}{\0}{+}{+}
\,\Bigr] \,,
  \phantom{\frac{1}{2}}
\nonumber \\
W_{LT}^{LT}(\phi_S,\phi,\varphi) &=
\sin(\phi-\phi_S)\, \Bigl[\,
  - \sin(\phi + \varphi)\, \sqrt{\epsilon (1-\epsilon)}\ms
    \re\bigl( \rn{\0}{+}{\0}{+} - \rn{-}{\0}{\0}{+} \bigr)
  \phantom{\frac{1}{2}}
\nonumber \\
& \qquad{}
  + \sin\varphi\, \sqrt{1 - \epsilon^2}\,
    \re\bigl( \rn{\0}{+}{+}{+} - \rn{-}{\0}{+}{+} \bigr)
  \phantom{\frac{1}{2}}
\nonumber \\
& \qquad{}
  + \sin(\phi - \varphi)\, \sqrt{\epsilon (1-\epsilon)}\ms
    \re\bigl( \rn{\0}{-}{\0}{+} - \rn{+}{\0}{\0}{+} \bigr)
\,\Bigr]
  \phantom{\frac{1}{2}}
\nonumber \\
& {}
+ \cos(\phi-\phi_S)\, \Bigl[\,
  \cos(\phi + \varphi)\, \sqrt{\epsilon (1-\epsilon)}\ms
  \re\bigl( \rs{\0}{+}{\0}{+} - \rs{-}{\0}{\0}{+} \bigr)
  \phantom{\frac{1}{2}}
\nonumber \\
& \qquad{}
  - \cos\varphi\, \sqrt{1 - \epsilon^2}\,
    \re\bigl( \rs{\0}{+}{+}{+} - \rs{-}{\0}{+}{+} \bigr)
  \phantom{\frac{1}{2}}
\nonumber \\
& \qquad{}
  - \cos(\phi - \varphi)\, \sqrt{\epsilon (1-\epsilon)}\ms
    \re\bigl( \rs{\0}{-}{\0}{+} - \rs{+}{\0}{\0}{+} \bigr)
\,\Bigr] \,,
  \phantom{\frac{1}{2}}
\nonumber \\
W_{LT}^{TT}(\phi_S,\phi,\varphi) &=
\sin(\phi-\phi_S)\,
  \phantom{\frac{1}{2}}
\nonumber \\
& \quad\times \Bigl[\,
  \sin\phi\, \sqrt{\epsilon (1-\epsilon)}\ms
    \re\bigl( \rn{+}{+}{\0}{+} + \rn{-}{-}{\0}{+} \bigr)
  - \sin(\phi + 2\varphi)\, \sqrt{\epsilon (1-\epsilon)}\ms
    \re\rn{-}{+}{\0}{+}
  \phantom{\frac{1}{2}}
\nonumber \\
& \qquad{}
  + \sin(2\varphi)\, \sqrt{1 - \epsilon^2}\, \re\rn{-}{+}{+}{+}
  \phantom{\frac{1}{2}}
\nonumber \\
& \qquad{}
  - \sin(\phi - 2\varphi)\, \sqrt{\epsilon (1-\epsilon)}\ms
    \re\rn{+}{-}{\0}{+}
\,\Bigr]
  \phantom{\frac{1}{2}}
\nonumber \\
& {}
+ \cos(\phi-\phi_S)\, \Bigl[\,
  \sqrt{1 - \epsilon^2}\; \tfrac{1}{2}\ms
  \bigl( \rs{+}{+}{+}{+} + \rs{-}{-}{+}{+} \bigr)
  \phantom{\frac{1}{2}}
\displaybreak[2]
\nonumber \\
& \qquad{}
  - \cos\phi\, \sqrt{\epsilon (1-\epsilon)}\ms
    \re\bigl( \rs{+}{+}{\0}{+} + \rs{-}{-}{\0}{+} \bigr)
  + \cos(\phi + 2\varphi)\, \sqrt{\epsilon (1-\epsilon)}\ms
    \re\rs{-}{+}{\0}{+}
  \phantom{\frac{1}{2}}
\nonumber \\
& \qquad{}
  - \cos(2\varphi)\, \sqrt{1 - \epsilon^2}\, \re\rs{-}{+}{+}{+}
  \phantom{\frac{1}{2}}
\nonumber \\
& \qquad{}
  + \cos(\phi - 2\varphi)\, \sqrt{\epsilon (1-\epsilon)}\ms
    \re\rs{+}{-}{\0}{+}
\,\Bigr]
  \phantom{\frac{1}{2}}
\end{align}
for beam polarization.  With obvious adjustments, the general
structure discussed in points 1 to 3 above is found again for a
transverse target.  Note that the terms $\ru{\0}{\0}{+}{+} + \epsilon
\ru{\0}{\0}{\0}{\0}$ and $\ru{+}{+}{+}{+} + \ru{-}{-}{+}{+} +
2\epsilon \ru{+}{+}{\0}{\0}$ in the unpolarized coefficients
$W_{UU}^{LL}$ and $W_{UU}^{TT}$ add up to $1$ according to
\eqref{norm-rel}, whereas their counterparts $\im\bigl(
\rn{\0}{\0}{+}{+} + \epsilon \rn{\0}{\0}{\0}{\0} \bigr)$ and
$\im\bigl( \rn{+}{+}{+}{+} + \rn{-}{-}{+}{+} + 2\epsilon
\rn{+}{+}{\0}{\0} \bigr)$ in $W_{UT}^{LL}$ and $W_{UT}^{TT}$ are
independent quantities.  To keep the close similarity between the two
cases we have not used \eqref{norm-rel} to simplify \eqref{WUU}.

Since there are two independent transverse polarizations relative to
the hadron plane (normal and sideways) we have a rather large number
of terms with different angular dependence in \eqref{WUT} and
\eqref{WLT}.  The single spin asymmetry $W_{UT}$ contains 16 terms
with $\im\rn{}{}{}{}$ and 14 terms with $\im\rs{}{}{}{}$, whereas the
double spin asymmetry $W_{LT}$ contains 8 terms with $\re\rn{}{}{}{}$
and 10 terms with $\re\rs{}{}{}{}$.  Table~\ref{tab:count} lists the
number of independent linear combinations of spin density matrix
elements describing the angular distribution for the different
combinations of beam and target spin.  For reasons discussed in
Section~\ref{sec:nat-par} it is useful to consider the spin density
matrices $\rn{}{}{}{}$ and $\rs{}{}{}{}$ separately.  It is then
natural to work in the basis of angular functions given by the product
of $\sin(\phi-\phi_S)$ or $\cos(\phi-\phi_S)$ with $\sin(k\phi +
m\varphi)$ or $\cos(k\phi + m\varphi)$.  With the replacement rules
\eqref{repl-rules-lu} and \eqref{repl-rules} we obtain the
combinations
\begin{align}
  \sin(\phi-\phi_S) \cos(k\phi + m\varphi) \ms\im\rn{}{}{}{} &
+ \cos(\phi-\phi_S) \sin(k\phi + m\varphi) \ms\im\rs{}{}{}{} \,,
\nonumber \\
- \sin(\phi-\phi_S) \sin(k\phi + m\varphi) \ms\re\rn{}{}{}{} &
+ \cos(\phi-\phi_S) \cos(k\phi + m\varphi) \ms\re\rs{}{}{}{}
\end{align}
in $W_{UT}$ and $W_{LT}$, respectively.

\TABLE[t]{
\caption{\label{tab:count} Number of linear combinations of spin
  density matrix elements describing the angular distribution of the
  cross section \protect\eqref{X-sect}.  The number of independent
  combinations for $\re\ru{}{}{}{}$ is one less than for
  $\im\rn{}{}{}{}$ because of the relation \protect\eqref{norm-rel}.}
\begin{tabular}{cccccccc} \hline\hline
\multicolumn{4}{c}{unpolarized beam} &
\multicolumn{4}{c}{polarized beam} \\
$W_{UU}$ & $W_{UL}$ & \multicolumn{2}{c}{$W_{UT}$} &
$W_{LU}$ & $W_{LL}$ & \multicolumn{2}{c}{$W_{LT}$} \\
~$\re\ru{}{}{}{}$~ & ~$\im\rl{}{}{}{}$~ &
 $\im\rn{}{}{}{}$  &  $\im\rs{}{}{}{}$  & 
~$\im\ru{}{}{}{}$~ & ~$\re\rl{}{}{}{}$~ &
 $\re\rn{}{}{}{}$  &  $\re\rs{}{}{}{}$  \\
15 & 14 & 16 & 14 &
 8 & 10 &  8 & 10 \\ \hline\hline
\end{tabular}
}


We conclude this section by giving the relation between our spin
density matrix elements for an unpolarized target and those in the
classical work \cite{Schilling:1973ag} of Schilling and Wolf.  We have
\begin{align}
  \label{schilling-re}
\ru{\0}{\0}{+}{+} + \epsilon \ru{\0}{\0}{\0}{\0} &= r^{04}_{00} \,,
\nonumber \\
\re \bigl( \ru{\0}{+}{\0}{+} - \ru{-}{\0}{\0}{+} \bigr)
 &= \sqrt{2}\, \bigl( \im r^6_{10} - \re r^5_{10} \bigr) \,,
\nonumber \\
\ru{+}{+}{+}{+} + \ru{-}{-}{+}{+} + 2\epsilon \ru{+}{+}{\0}{\0}
 &= 1 - r^{04}_{00} \,,
\nonumber \\
\ru{-}{+}{-}{+} &= r^1_{1-1} - \im r^2_{1-1} \,,
\displaybreak[2]
\nonumber \\
\re \ru{\0}{\0}{\0}{+} &=  - r^5_{00} /\sqrt{2} \,,
\nonumber \\
\re \bigl( \ru{\0}{+}{+}{+} - \ru{-}{\0}{+}{+}
  + 2\epsilon \ru{\0}{+}{\0}{\0} \bigr) &= 2 \re r^{04}_{10} \,,
\nonumber \\
\re \ru{\0}{+}{-}{+} &= \re r^1_{10} - \im r^2_{10} \,,
\nonumber \\
\re \bigl( \ru{\0}{-}{\0}{+} - \ru{+}{\0}{\0}{+} \bigr)
 &= \sqrt{2}\, \bigl( \im r^6_{10} + \re r^5_{10} \bigr) \,,
\nonumber \\
\re \bigl( \ru{-}{+}{+}{+} + \epsilon \ru{-}{+}{\0}{\0} \bigr)
 &= r^{04}_{1-1} \,,
\nonumber \\
\re \ru{+}{+}{-}{+} &= r^1_{11} \,,
\nonumber \\
\re \bigl( \ru{+}{+}{\0}{+} + \ru{-}{-}{\0}{+} \bigr)
 &= - \sqrt{2}\, r^5_{11} \,,
\nonumber \\
\re \ru{-}{+}{\0}{+}
 &= \bigl( \im r^6_{1-1} - r^5_{1-1} \bigr) /\sqrt{2} \,,
\nonumber \\
\ru{\0}{\0}{-}{+} &= r^1_{00} \,,
\nonumber \\
\re \ru{+}{\0}{-}{+} &= \re r^1_{10} + \im r^2_{10} \,,
\nonumber \\
\re \ru{+}{-}{\0}{+}
 &= - \bigl( \im r^6_{1-1} + r^5_{1-1} \bigr) /\sqrt{2} \,,
\nonumber \\
\ru{+}{-}{-}{+} &= r^1_{1-1} + \im r^2_{1-1}
\\
%
\intertext{and}
  \label{schilling-im}
\im \bigl( \ru{\0}{+}{\0}{+} - \ru{-}{\0}{\0}{+} \bigr)
 &= \sqrt{2}\, \bigl( \im r^7_{10} + \re r^8_{10} \bigr) \,,
\nonumber \\
\im \ru{\0}{\0}{\0}{+} &= r^8_{00} /\sqrt{2} \,,
\nonumber \\
\im \bigl( \ru{\0}{+}{+}{+} - \ru{-}{\0}{+}{+} \bigr)
 &= - 2 \im r^{3}_{10} \,,
\nonumber \\
\im \bigl( \ru{\0}{-}{\0}{+} - \ru{+}{\0}{\0}{+} \bigr)
 &= \sqrt{2}\, \bigl( \im r^7_{10} - \re r^8_{10} \bigr) \,,
\nonumber \\
\im \ru{-}{+}{+}{+} &= - \im r^3_{1-1} \,,
\nonumber \\
\im \bigl( \ru{+}{+}{\0}{+} + \ru{-}{-}{\0}{+} \bigr)
 &= \sqrt{2}\, r^8_{11} \,,
\nonumber \\
\im \ru{-}{+}{\0}{+}
 &= \bigl( \im r^7_{1-1} + r^8_{1-1} \bigr) /\sqrt{2} \,,
\nonumber \\
\im \ru{+}{-}{\0}{+}
 &= - \bigl( \im r^7_{1-1} - r^8_{1-1} \bigr) /\sqrt{2} \,.
\end{align}
The lower indices in the matrix elements of Schilling and Wolf refer
to the $\rho$ helicity and correspond to the upper indices of
$\ru{}{}{}{}$ in our notation.  Their upper indices correspond to a
representation of the virtual photon spin density matrix which refers
partly to circular and partly to linear polarization, whereas we use
the helicity basis for the photon throughout.  The consequences of
approximate $s$-channel helicity conservation are more explicit in our
notation: the relation $\im r_{10}^6 \approx - \re r_{10}^5$ for
instance corresponds to $\bigl|\ms \re \bigl( \ru{\0}{+}{\0}{+} -
\ru{-}{\0}{\0}{+} \bigr) \bigr| \gg \bigl|\ms \re \bigl(
\ru{\0}{-}{\0}{+} - \ru{+}{\0}{\0}{+} \bigr) \bigr|$.  Notice also
that the simple relation between single-spin asymmetries and imaginary
parts of spin density matrix elements discussed in point 3 above holds
in the helicity basis but not for linear polarization.

We note that our phase convention \eqref{photon-eps} for the helicity
states of the virtual photon differs from the one in
\cite{Schilling:1973ag} by a relative minus sign between transverse
and longitudinal polarization, and that our normalization factors
$N_T$ and $N_L$ in \eqref{norm-def} differ from those in
\cite{Schilling:1973ag} by a factor of two.  The combinations of
helicity amplitudes corresponding to the spin density matrix elements
in \eqref{schilling-re} and \eqref{schilling-im} should be compared
according to
\begin{equation}
  \label{schilling-T}
\frac{1}{2}\, \Biggr[ 
  \frac{1}{N_T + \epsilon N_L}\, \sum_{\sigma\lambda}
     T^{\ms\nu \sigma}_{\mu \lambda}\, \bigl (
     T^{\ms\nu'\sigma}_{\mu'\lambda} \bigr)^*
  \Biggr]_{\ms\text{here}}^{}
= \eta_{\ms \mu\mu'}\, \Biggr[
  \frac{1}{N_T + \epsilon N_L}\, \sum_{\sigma\lambda}
     T_{\nu\sigma,\ms \mu\lambda}^{\phantom{*}}\,
     T_{\nu'\sigma,\ms \mu'\lambda}^{*}
  \Biggr]_{\ms\text{\protect\cite{Schilling:1973ag}}} \,\,,
\end{equation}
where $\eta_{\0 \pm}^{} = \eta_{\pm \0}^{} = -1$ for the interference
of transverse and longitudinal photon polarization, and $\eta_{\ms
\mu\mu'} = +1$ in all other cases.\footnote{%
The correspondence in \protect\eqref{schilling-re} to
\protect\eqref{schilling-T} is obtained from comparing our results
\eqref{WUU} and \eqref{WLU} for the angular distribution with the ones
in eqs.~(92) and (92a) of \protect\cite{Schilling:1973ag}, together
with the relation between spin density matrix elements and helicity
amplitudes specified in eq.~(91) and Appendix~A of
\protect\cite{Schilling:1973ag}.  We have not found an explicit
specification of the phase convention for the virtual photon
polarizations used in \protect\cite{Schilling:1973ag}.}


\section{Natural and unnatural parity}
\label{sec:nat-par}

The exclusive process $\gamma^* p\to \rho\ms p$ is described by
eighteen independent helicity amplitudes, and we have already used
approximate $s$-channel helicity conservation to establish a hierarchy
among these amplitudes and the spin density matrix elements
constructed from them.  A further dynamical criterion to order these
quantities is given by natural and unnatural parity exchange, which we
shall now discuss.

Following \cite{Schilling:1973ag} we define amplitudes $N$ for natural
and $U$ unnatural parity exchange as linear combinations
\begin{align}
N^{\nu\sigma}_{\mu\lambda}
 &= \half\ms \bigl[\ms T^{\nu\sigma}_{\mu\lambda}
    + (-1)^{\nu-\mu}\, T^{-\nu\sigma}_{-\mu\lambda} \ms\bigr]
  = \half\ms \bigl[\ms T^{\nu\sigma}_{\mu\lambda}
    + (-1)^{\lambda-\sigma}\, T^{\ms\nu\ms-\sigma}_{\mu\ms-\lambda}
    \ms\bigr] \,,
\nonumber \\
U^{\nu \sigma}_{\mu \lambda}
 &= \half\ms \bigl[\ms T^{\ms\nu \sigma}_{\mu \lambda}
    - (-1)^{\nu-\mu}\, T^{-\nu \sigma}_{-\mu \lambda} \ms\bigr]
  = \half\ms \bigl[\ms T^{\ms\nu \sigma}_{\mu \lambda}
    - (-1)^{\lambda-\sigma}\, T^{\ms\nu\ms-\sigma}_{\mu\ms-\lambda}
    \ms\bigr] \,.
\end{align}
With respect to the photon and meson helicity, the amplitudes $N$ have
the same symmetry behavior as the amplitudes for $\gamma^* t\to
\rho\ms t$ on a spin-zero target $t$, whereas the corresponding
relation for the amplitudes $U$ has an additional minus sign,
\begin{align}
  \label{un-parity}
N^{-\nu\sigma}_{-\mu\lambda} 
  &= (-1)^{\nu-\mu}\, N^{\nu\sigma}_{\mu\lambda} \,,
&
U^{-\nu\sigma}_{-\mu\lambda} 
  &= - (-1)^{\nu-\mu}\, U^{\nu\sigma}_{\mu\lambda} \,.
\end{align}
For the proton helicity we have relations $N^{\nu+}_{\mu+} =
N^{\nu-}_{\mu-}$ and $N^{\nu+}_{\mu-} = -N^{\nu-}_{\mu+}$ for natural
parity exchange, compared to $U^{\nu+}_{\mu+} = -U^{\nu-}_{\mu-}$ and
$U^{\nu+}_{\mu-} = U^{\nu-}_{\mu+}$ for unnatural parity exchange.
This symmetry behavior immediately implies that in a dynamical
description using generalized parton distributions, amplitudes $N$ go
with distributions $H$ and $E$, whereas amplitudes $U$ go with
distributions $\tilde{H}$ and $\tilde{E}$.  This is explicitly borne
out in the calculation of \cite{Goloskokov:2005sd}.  Since $U^{\0
\sigma}_{\0 \lambda} = 0$ according to \eqref{un-parity}, unnatural
parity exchange amplitudes are power suppressed at large $Q^2$ and the
leading-twist factorization theorem \cite{Collins:1996fb} only applies
to the natural parity exchange amplitudes $N^{\0 \sigma}_{\0
\lambda}$.
We remark that in the context of low-energy dynamics $t$-channel
exchange of a pion plays a prominent role for unnatural parity
exchange amplitudes, see e.g.~\cite{Fraas:1974vx}.  This has a natural
counterpart in the framework of generalized parton distributions,
where pion exchange gives an essential contribution to the
distribution $\tilde{E}$ in the isovector channel
\cite{Mankiewicz:1998kg,Goeke:2001tz,Ando:2006sk}.

For the spin density matrix elements one readily finds
\begin{align}
  \label{un-sdme}
\ru{\nu}{\nu'}{\mu}{\mu'} &= (N_T + \epsilon N_L)^{-1}\,
  \sum_\sigma\, \Bigl[\,
    N^{\nu \sigma}_{\mu +}\,
    \bigl( N^{\nu'\sigma}_{\mu' +} \bigr)^*
  + U^{\nu \sigma}_{\mu +}\,
    \bigl( U^{\nu'\sigma}_{\mu' +} \bigr)^* \,\Bigr] \,,
\nonumber \\
\rl{\nu}{\nu'}{\mu}{\mu'} &= (N_T + \epsilon N_L)^{-1}\,
  \sum_\sigma\, \Bigl[\,
    N^{\nu \sigma}_{\mu +}\,
    \bigl( U^{\ms\nu'\sigma}_{\mu' +} \bigr)^*
  + U^{\nu \sigma}_{\mu +}\,
    \bigl( N^{\ms\nu'\sigma}_{\mu' +} \bigr)^* \,\Bigr] \,,
\nonumber \\
\rs{\nu}{\nu'}{\mu}{\mu'} &= (N_T + \epsilon N_L)^{-1}\,
  \sum_\sigma\, \Bigl[\,
    N^{\nu \sigma}_{\mu +}\,
    \bigl( U^{\ms\nu'\sigma}_{\mu' -} \bigr)^*
  + U^{\nu \sigma}_{\mu +}\,
    \bigl( N^{\ms\nu'\sigma}_{\mu' -} \bigr)^* \,\Bigr] \,,
\nonumber \\
\rn{\nu}{\nu'}{\mu}{\mu'} &= (N_T + \epsilon N_L)^{-1}\,
  \sum_\sigma\, \Bigl[\,
    N^{\nu \sigma}_{\mu +}\,
    \bigl( N^{\ms\nu'\sigma}_{\mu' -} \bigr)^*
  + U^{\nu \sigma}_{\mu +}\,
    \bigl( U^{\ms\nu'\sigma}_{\mu' -} \bigr)^* \,\Bigr] \,.
\end{align}
The matrix elements $\ru{}{}{}{}$ and $\rn{}{}{}{}$ hence involve a
product of two natural parity exchange amplitudes plus a product of
two amplitudes for unnatural parity exchange, whereas $\rl{}{}{}{}$
and $\rs{}{}{}{}$ involve the interference between natural and
unnatural parity exchange \cite{Fraas:1974vx}.  To the extent that
amplitudes $U$ are smaller than their counterparts $N$, one can thus
expect that matrix elements $\rl{}{}{}{}$ and $\rs{}{}{}{}$ are small
compared with $\ru{}{}{}{}$ and $\rn{}{}{}{}$ for equal helicity
indices.  Exceptions to this guideline are possible since products
$N^{\nu \sigma}_{\mu +}\, \bigl( N^{\nu'\sigma}_{\mu' +} \bigr)^*$ or
$N^{\nu \sigma}_{\mu -}\, \bigl( N^{\ms\nu'\sigma}_{\mu' +} \bigr)^*$
may have a small real or imaginary part due to the relative phase
between the two amplitudes.  If amplitudes $U$ are smaller than $N$,
one can furthermore neglect the terms
\begin{align}
\ruu{\nu}{\nu'}{\mu}{\mu'} &= 
(N_T + \epsilon N_L)^{-1}\, \sum_\sigma\,
    U^{\nu \sigma}_{\mu +}\,
    \bigl( U^{\nu'\sigma}_{\mu' +} \bigr)^* \,,
\nonumber \\
\rnu{\nu}{\nu'}{\mu}{\mu'} &= 
(N_T + \epsilon N_L)^{-1}\, \sum_\sigma\,
    U^{\nu \sigma}_{\mu +}\,
    \bigl( U^{\nu'\sigma}_{\mu' -} \bigr)^*
\end{align}
involving unnatural parity exchange in the matrix elements
$\ru{}{}{}{}$ and $\rn{}{}{}{}$.
Using the relations 
\begin{equation}
(-1)^{\nu-\mu}\,   \ru{-\nu\ms}{\nu'}{-\mu\ms}{\mu'} =
\ru{\nu}{\nu'}{\mu}{\mu'} - 2 \ruu{\nu}{\nu'}{\mu}{\mu'}
\end{equation}
following from \eqref{un-parity} and \eqref{un-sdme}, we have in
particular
\begin{align}
  \label{un-equal}
-\ru{\0}{+}{-}{+} &= \ru{\0}{+}{+}{+} - 2 \ruu{\0}{+}{+}{+} \,,
&
  \ru{-}{+}{-}{+} &= \ru{+}{+}{+}{+} - 2 \ruu{+}{+}{+}{+} \,,
\nonumber \\[0.2em]
-\ru{-}{+}{\0}{+} &= \ru{+}{+}{\0}{+} - 2 \ruu{+}{+}{\0}{+} \,,
&
  \ru{+}{+}{-}{+} &= \ru{-}{+}{+}{+} - 2 \ruu{-}{+}{+}{+} \,.
\end{align}
This allows us to rewrite
\begin{align}
  \label{un-WU}
W_{UU}^{LT} &=
    - \cos\varphi\,
      \re\bigl( \ru{\0}{+}{+}{+} - \ru{-}{\0}{+}{+}
             + 2\epsilon \ru{\0}{+}{\0}{\0} \bigr)
    - \cos(2\phi + \varphi)\, \epsilon
      \re\bigl( \ru{\0}{+}{+}{+} - 2 \ruu{\0}{+}{+}{+} \bigr)
    \phantom{\frac{1}{2}}
\nonumber \\
& \quad{}
    + \ldots \cos(\phi + \varphi)
    + \ldots \cos(\phi - \varphi)
    + \ldots \cos(2\phi - \varphi) \,,
    \phantom{\frac{1}{2}}
\nonumber \\
W_{UU}^{TT} &=
    \tfrac{1}{2}\ms \bigl( \ru{+}{+}{+}{+} + \ru{-}{-}{+}{+}
                         + 2\epsilon \ru{+}{+}{\0}{\0} \bigr)
    + \tfrac{1}{2} \cos(2\phi + 2\varphi)\, \epsilon
      \bigl( \ru{+}{+}{+}{+} - 2 \ruu{+}{+}{+}{+} \bigr)
    \phantom{\frac{1}{2}}
\nonumber \\
& \quad{}
    - \cos\phi\, \sqrt{\epsilon (1+\epsilon)}\ms
      \re\bigl( \ru{+}{+}{\0}{+} + \ru{-}{-}{\0}{+} \bigr)
    - \cos(\phi + 2\varphi)\, \sqrt{\epsilon (1+\epsilon)}\ms
      \re\bigl( \ru{+}{+}{\0}{+} - 2 \ruu{+}{+}{\0}{+} \bigr)
    \phantom{\frac{1}{2}}
\nonumber \\
& \quad{}
    - \cos(2\varphi)\,
      \re\bigl( \ru{-}{+}{+}{+} + \epsilon \ru{-}{+}{\0}{\0} \bigr)
    - \cos(2\phi)\, \epsilon
      \re\bigl( \ru{-}{+}{+}{+} - 2 \ruu{-}{+}{+}{+} \bigr)
    \phantom{\frac{1}{2}}
\nonumber \\
& \quad{}
    + \ldots \cos(\phi - 2\varphi)
    + \ldots \cos(2\phi - 2\varphi) \,,
    \phantom{\frac{1}{2}}
\displaybreak[2]
\nonumber \\
W_{LU}^{TT} &=
    - \sin\phi\, \sqrt{\epsilon (1-\epsilon)}\ms
      \im\bigl( \ru{+}{+}{\0}{+} + \ru{-}{-}{\0}{+} \bigr)
    - \sin(\phi + 2\varphi)\, \sqrt{\epsilon (1-\epsilon)}\ms
      \im\bigl( \ru{+}{+}{\0}{+} - 2 \ruu{+}{+}{\0}{+} \bigr)
    \phantom{\frac{1}{2}}
\nonumber \\
& \quad{}
    + \ldots \sin(2\varphi)
    + \ldots \sin(\phi - 2\varphi) \,,
    \phantom{\frac{1}{2}}
\end{align}
where terms indicated by $\ldots$ are the same as in the original
expressions \eqref{WUU} and \eqref{WLU} and have not been repeated for
brevity.  We see that the coefficients of adjacent terms in
\eqref{un-WU} will be approximately equal to the extent that unnatural
parity exchange is suppressed and $s$-channel helicity approximately
conserved.  This can be tested experimentally by measuring the angular
distribution of the final-state particles.

The relations \eqref{un-equal} and their counterparts for other index
combinations can also be used to approximately isolate spin density
matrix elements of particular interest.  Consider as an example the
leading-twist matrix element $\ru{\0}{\0}{\0}{\0}$, which in the
angular distribution appears only in the combination
$\ru{\0}{\0}{+}{+} + \epsilon \ru{\0}{\0}{\0}{\0}$, i.e.\ together
with a matrix element that should be suppressed since it does not
conserve $s$-channel helicity.  If unnatural parity exchange is
strongly suppressed, an even better approximation for
$\ru{\0}{\0}{\0}{\0}$ can be obtained from the linear combination
\begin{align}
\epsilon \ru{\0}{\0}{\0}{\0} + 2 \ruu{\0}{\0}{+}{+}
 &= \bigl( \ru{\0}{\0}{+}{+} + \epsilon \ru{\0}{\0}{\0}{\0} \bigr)
    + \ru{\0}{\0}{-}{+} \,,
\end{align}
whose r.h.s.\ can be extracted from the angular distribution.
Similarly, one can approximately isolate the matrix element
$\re\ru{\0}{+}{\0}{\0}$ in the combination
\begin{align}
\epsilon \re\ru{\0}{+}{\0}{\0}
       + \re \bigl( \ruu{\0}{+}{+}{+} - \ruu{-}{\0}{+}{+} \bigr)
 &= \half \Bigl[ \re \bigl( \ru{\0}{+}{+}{+} - \ru{-}{\0}{+}{+}
              + 2\epsilon \ru{\0}{+}{\0}{\0} \bigr)
              + \re\ru{\0}{+}{-}{+} + \re\ru{+}{\0}{-}{+} \Bigr] \,.
\end{align}
Conversely, one can extract from the angular distribution the linear
combinations
\begin{align}
\ruu{+}{+}{+}{+} + \ruu{+}{+}{-}{-} 
    + 2\epsilon \ruu{+}{+}{\0}{\0}
    - 2\re\ruu{+}{+}{-}{+}
 &= \half\ms \bigl(\ru{+}{+}{+}{+} + \ru{-}{-}{+}{+}
          + 2\epsilon \ru{+}{+}{\0}{\0} \bigr)
    - \half\ms \ru{-}{+}{-}{+} - \half\ms \ru{+}{-}{-}{+}
\nonumber \\[0.3em]
 & \quad
    + \re \bigl( \ru{-}{+}{+}{+} + \epsilon \ru{-}{+}{\0}{\0} \bigr)
    - \re\ru{+}{+}{-}{+} \,,
\nonumber \\[0.3em]
\ruu{+}{+}{\0}{+} + \ruu{-}{-}{\0}{+}
 &= \half \bigl( \ru{+}{+}{\0}{+} + \ru{-}{-}{\0}{+} \bigr)
    + \half\ms \ru{-}{+}{\0}{+} + \half\ms \ru{+}{-}{\0}{+} \,,
\end{align}
which only involve unnatural parity exchange.  In a dynamical approach
based on generalized parton distributions, these combinations are
interesting because they isolate the polarized distributions
$\tilde{H}$ and $\tilde{E}$ and in particular involve these
distributions for gluons, which are very hard to access in any other
process.\footnote{%
In contrast to their quark counterparts, $\tilde{H}^g$ and
$\tilde{E}^g$ do not appear in pseudoscalar meson production at
leading twist and leading order in $\alpha_s$, see e.g. Section~5.1.1
of \protect\cite{Diehl:2003ny}.}
The price to pay for this is that the corresponding amplitudes are
power suppressed and cannot be calculated with the theoretical rigor
provided by the leading-twist factorization theorem.  On the other
hand, phenomenological analysis indicates that a quantitative
description of meson production at $Q^2$ of a few $\gev^2$ requires
the inclusion of power-suppressed effects also for the leading matrix
element $\ru{\0}{\0}{\0}{\0}$.

The discussion of the matrix elements for transverse target
polarization normal to the hadron plane proceeds in full analogy to
the unpolarized case.  With
\begin{equation}
(-1)^{\nu-\mu}\,   \rn{-\nu\ms}{\nu'}{-\mu\ms}{\mu'} =
\rn{\nu}{\nu'}{\mu}{\mu'} - 2 \rnu{\nu}{\nu'}{\mu}{\mu'}
\end{equation}
we have
\begin{align}
-\rn{\0}{+}{-}{+} &= \rn{\0}{+}{+}{+} - 2 \rnu{\0}{+}{+}{+} \,,
&
  \rn{-}{+}{-}{+} &= \rn{+}{+}{+}{+} - 2 \rnu{+}{+}{+}{+} \,,
\nonumber \\[0.2em]
-\rn{-}{+}{\0}{+} &= \rn{+}{+}{\0}{+} - 2 \rnu{+}{+}{\0}{+} \,,
&
  \rn{+}{+}{-}{+} &= \rn{-}{+}{+}{+} - 2 \rnu{-}{+}{+}{+}
\end{align}
and can write
\begin{align}
W_{UT}^{LT} &=
\cos(\phi-\phi_S)\, \Bigl[\; \ldots \;\Bigr] + \sin(\phi-\phi_S)\,
  \phantom{\frac{1}{2}}
\nonumber \\
& \quad \times \Bigl[\,
  - \cos\varphi\,
    \im\bigl( \rn{\0}{+}{+}{+} - \rn{-}{\0}{+}{+}
          + 2\epsilon \rn{\0}{+}{\0}{\0} \bigr)
  - \cos(2\phi + \varphi)\, \epsilon
    \im\bigl( \rn{\0}{+}{+}{+} - 2\rnu{\0}{+}{+}{+} \bigr)
  \phantom{\frac{1}{2}}
\nonumber \\
& \qquad
  + \ldots \cos(\phi + \varphi)
  + \ldots \cos(\phi - \varphi)
  + \ldots \cos(2\phi - \varphi)
\,\Bigr] \,,
\nonumber \\
W_{UT}^{TT} &=
\cos(\phi-\phi_S)\, \Bigl[\; \ldots \;\Bigr] + \sin(\phi-\phi_S)\, 
  \phantom{\frac{1}{2}}
\nonumber \\
& \quad \times \Bigl[\,
  \tfrac{1}{2} \im\bigl( \rn{+}{+}{+}{+} + \rn{-}{-}{+}{+}
                      + 2\epsilon \rn{+}{+}{\0}{\0} \bigr)
  + \tfrac{1}{2} \cos(2\phi + 2\varphi)\, \epsilon
    \im\bigl( \rn{+}{+}{+}{+} - 2\rnu{+}{+}{+}{+} \bigr)
  \phantom{\frac{1}{2}}
\nonumber \\
& \qquad
  - \cos\phi\, \sqrt{\epsilon (1+\epsilon)}\ms
    \im\bigl( \rn{+}{+}{\0}{+} + \rn{-}{-}{\0}{+} \bigr)
  - \cos(\phi + 2\varphi)\, \sqrt{\epsilon (1+\epsilon)}\ms
    \im\bigl( \rn{+}{+}{\0}{+} -2\rnu{+}{+}{\0}{+} \bigr)
  \phantom{\frac{1}{2}}
\nonumber \\
& \qquad
  - \cos(2\varphi)\,
    \im\bigl( \rn{-}{+}{+}{+} + \epsilon \rn{-}{+}{\0}{\0} \bigr)
  - \cos(2\phi)\, \epsilon 
    \im\bigl( \rn{-}{+}{+}{+} -2\rnu{-}{+}{+}{+} \bigr)
  \phantom{\frac{1}{2}}
\nonumber \\
& \qquad
  + \ldots \cos(\phi - 2\varphi)
  + \ldots \cos(2\phi - 2\varphi)
\,\Bigr] \,,
\nonumber \\
W_{LT}^{TT} &=
\cos(\phi-\phi_S)\, \Bigl[\; \ldots \;\Bigr] + \sin(\phi-\phi_S)\,
  \phantom{\frac{1}{2}}
\nonumber \\
& \quad \times \Bigl[\,
  \sin\phi\, \sqrt{\epsilon (1-\epsilon)}\ms
    \re\bigl( \rn{+}{+}{\0}{+} + \rn{-}{-}{\0}{+} \bigr)
  + \sin(\phi + 2\varphi)\, \sqrt{\epsilon (1-\epsilon)}\ms
    \re \bigl( \rn{+}{+}{\0}{+} - 2\rnu{+}{+}{\0}{+} \bigr)
  \phantom{\frac{1}{2}}
\nonumber \\
& \qquad
  + \ldots \sin(2\varphi)
  + \ldots \sin(\phi - 2\varphi)
\,\Bigr] \,,
  \phantom{\frac{1}{2}}
\end{align}
where terms denoted by $\ldots$ are as in the original expressions
\eqref{WUT} and \eqref{WLT}.  Again, the coefficients of adjacent
terms should be approximately equal to the extent that unnatural
parity exchange is suppressed and $s$-channel helicity approximately
conserved.  The matrix elements $\im\rn{\0}{\0}{\0}{\0}$ and
$\im\rn{\0}{+}{\0}{\0}$ can be approximately isolated in
\begin{align}
  \label{un-n1}
\epsilon \im\rn{\0}{\0}{\0}{\0} + 2 \im\rnu{\0}{\0}{+}{+}
 &= \im \bigl( \rn{\0}{\0}{+}{+} + \epsilon \rn{\0}{\0}{\0}{\0} \bigr)
    + \im\rn{\0}{\0}{-}{+}
\\[0.4em]
\intertext{and}
  \label{un-n2}
\epsilon \im\rn{\0}{+}{\0}{\0}
       + \im \bigl( \rnu{\0}{+}{+}{+} - \rnu{-}{\0}{+}{+} \bigr)
 &= \half \Bigl[ \im \bigl( \rn{\0}{+}{+}{+} - \rn{-}{\0}{+}{+}
              + 2\epsilon \rn{\0}{+}{\0}{\0} \bigr)
              + \im\rn{\0}{+}{-}{+} + \im\rn{+}{\0}{-}{+} \Bigr] .
\end{align}
In turn, the linear combinations
\begin{align}
\im \bigl( \rnu{+}{+}{+}{+} + \rnu{+}{+}{-}{-} 
        + 2\epsilon \rnu{+}{+}{\0}{\0}
        - 2\rnu{+}{+}{-}{+} \bigr)
 &= \half \im \bigl(\rn{+}{+}{+}{+} + \rn{-}{-}{+}{+}
          + 2\epsilon \rn{+}{+}{\0}{\0} \bigr)
    - \half \im\rn{-}{+}{-}{+} - \half \im\rn{+}{-}{-}{+}
\nonumber \\[0.3em]
 & \quad
    + \im \bigl( \rn{-}{+}{+}{+} + \epsilon \rn{-}{+}{\0}{\0} \bigr)
    - \im\rn{+}{+}{-}{+} \,,
\nonumber \\[0.3em]
\rnu{+}{+}{\0}{+} + \rnu{-}{-}{\0}{+}
 &= \half \bigl( \rn{+}{+}{\0}{+} + \rn{-}{-}{\0}{+} \bigr)
    + \half\ms \rn{-}{+}{\0}{+} + \half\ms \rn{+}{-}{\0}{+}
\end{align}
involve only unnatural parity exchange.


\section{Positivity constraints}
\label{sec:pos}

{}From the definition \eqref{sdm-def} of the spin-density matrix
elements one readily finds
\begin{align}
\sum_{\nu\mu\lambda}\, \sum_{\nu'\mu'\lambda'}
  c^{\ms\nu}_{\mu\lambda}\,
  \rho^{\ms\nu\nu'}_{\mu\mu', \lambda\lambda'}\,
  \bigl( c^{\ms\nu'}_{\mu'\lambda'} \bigr)^*
&= (N_T + \epsilon N_L)^{-1}\, \sum_{\sigma\rule{0pt}{1.2ex}}\,
   \Bigl| \sum_{\nu\mu\lambda}
   c^{\ms\nu}_{\mu\lambda}\,
   T^{\ms\nu \sigma}_{\mu \lambda} \,\Bigr|^2
\,\ge\, 0
\end{align}
for arbitrary complex numbers $c^{\nu}_{\mu\lambda}$.  Hence
$\rho^{\ms\nu\nu'}_{\mu\mu', \lambda\lambda'}$ is a positive
semidefinite matrix, with row indices specified by $\{ \nu\mu\lambda
\}$ and column indices by $\{ \nu'\mu'\lambda' \}$.  This implies
inequalities among the spin density matrix elements, which extend
those given e.g.\ in \cite{Diehl:2005pc,Arens:1996xw}.  We do not
attempt here to study the bounds following from positivity of the full
$18 \times 18$ matrix $\rho^{\ms\nu\nu'}_{\mu\mu', \lambda\lambda'}$,
which is quite unwieldy.  Instead, we consider the subset of matrix
elements conserving $s$-channel helicity for the photon-meson
transition and derive a number of simple inequalities, which may be
useful in practice.
Ordering the row and column indices as $\{+\!+\!+\}, \{\ms 0\, 0\ms
+\}, \{-\!-\!+\}, \{+\!+\!-\}, \{\ms 0\, 0\ms -\}, \{-\!-\!-\}$, we
have a positive semidefinite matrix $\mathrm{C}$, which can be written
in block form as
\begin{equation}
\mathrm{C} = 
\begin{pmatrix}
  \;\mathrm{A}_+\; & \mathrm{B}_{+} \\[0.2em]
  \mathrm{B}_{-} & \;\mathrm{A}_{-}\;
\end{pmatrix}
\end{equation}
with
\begin{align}
\mathrm{A}_{\eta} &= \begin{pmatrix}
  \ru{+}{+}{+}{+} + \eta\,\rl{+}{+}{+}{+} & 
  \bigl( \ru{\0}{+}{\0}{+} + \eta\,\rl{\0}{+}{\0}{+} \bigr)^* &
\; \ru{-}{+}{-}{+} - \eta\,\rl{-}{+}{-}{+} \;
\\[0.4em]
  \ru{\0}{+}{\0}{+} + \eta\,\rl{\0}{+}{\0}{+} &
  \ru{\0}{\0}{\0}{\0} &
  \ru{\0}{+}{\0}{+} - \eta\,\rl{\0}{+}{\0}{+}
\\[0.4em]
\; \ru{-}{+}{-}{+} + \eta\,\rl{-}{+}{-}{+} \; &
  \bigl( \ru{\0}{+}{\0}{+} - \eta\,\rl{\0}{+}{\0}{+} \bigr)^* &
  \ru{+}{+}{+}{+} - \eta\,\rl{+}{+}{+}{+}
\end{pmatrix}
\end{align}
and
\begin{align}
\mathrm{B}_{\eta} &= \begin{pmatrix}
  \rs{+}{+}{+}{+} + \eta\,\rn{+}{+}{+}{+} &
  \bigl( \rs{\0}{+}{\0}{+} - \eta\,\rn{\0}{+}{\0}{+} \bigr)^* &
\; -\rs{-}{+}{-}{+} + \eta\,\rn{-}{+}{-}{+} \;
\\[0.4em]
  \rs{\0}{+}{\0}{+} + \eta\,\rn{\0}{+}{\0}{+} &
  \eta\,\rn{\0}{\0}{\0}{\0} &
  -\rs{\0}{+}{\0}{+} + \eta\,\rn{\0}{+}{\0}{+}
\\[0.4em]
\; \rs{-}{+}{-}{+} + \eta\,\rn{-}{+}{-}{+} \; &
  - \bigl( \rs{\0}{+}{\0}{+} + \eta\,\rn{\0}{+}{\0}{+} \bigr)^* &
  -\rs{+}{+}{+}{+} + \eta\,\rn{+}{+}{+}{+}
\end{pmatrix} \,,
\end{align}
where $\eta=\pm 1$.  Concentrating first on the matrix elements for an
unpolarized or longitudinally polarized target, we find that the
matrix $\mathrm{A}_{\eta}$ has eigenvalues whose expressions are very
lengthy and therefore restrict our attention to $2\times 2$
submatrices.  The matrix obtained from the first and third rows and
columns of $\mathrm{A}_+$ has eigenvalues
\begin{equation}
\ru{+}{+}{+}{+} \pm \sqrt{
   \bigl( \ru{-}{+}{-}{+} \bigr)^2 
 + \bigl( \rl{+}{+}{+}{+} \bigr)^2 
 + \bigl( \im\rl{-}{+}{-}{+} \bigr)^2} \,,
\end{equation}
whose positivity implies a bound
\begin{equation}
  \label{l++bound}
\bigl( \rl{+}{+}{+}{+} \bigr)^2 
 + \bigl( \im\rl{-}{+}{-}{+} \bigr)^2
\le \bigl( \ru{+}{+}{+}{+} \bigr)^2 
    - \bigl( \ru{-}{+}{-}{+} \bigr)^2 \,.
\end{equation}
Similarly, the matrix obtained from the first and second and the
matrix obtained from the second and third rows and columns of
$\mathrm{A}_+$ have respective eigenvalues
\begin{align}
& \frac{1}{2}\ms \bigl( \ru{+}{+}{+}{+} + \rl{+}{+}{+}{+}
                      + \ru{\0}{\0}{\0}{\0} \bigr)
\pm \frac{1}{2}\ms
      \sqrt{ \bigl( \ru{+}{+}{+}{+} + \rl{+}{+}{+}{+}
                  - \ru{\0}{\0}{\0}{\0} \bigr)^2
    + 4\, \bigl| \ru{\0}{+}{\0}{+} + \rl{\0}{+}{\0}{+} \bigr|^2} \,,
\nonumber \\[0.2em]
& \frac{1}{2}\ms \bigl( \ru{+}{+}{+}{+} - \rl{+}{+}{+}{+}
                      + \ru{\0}{\0}{\0}{\0} \bigr)
\pm \frac{1}{2}\ms 
      \sqrt{ \bigl( \ru{+}{+}{+}{+} - \rl{+}{+}{+}{+}
                  - \ru{\0}{\0}{\0}{\0} \bigr)^2
    + 4\, \bigl| \ru{\0}{+}{\0}{+} - \rl{\0}{+}{\0}{+} \bigr|^2} \,,
\end{align}
whose positivity gives bounds
\begin{align}
  \bigl( \re\ru{\0}{+}{\0}{+} + \re\rl{\0}{+}{\0}{+} \bigr)^2
+ \bigl( \im\ru{\0}{+}{\0}{+} + \im\rl{\0}{+}{\0}{+} \bigr)^2
& \le \ru{\0}{\0}{\0}{\0}\,
    \bigl( \ru{+}{+}{+}{+} + \rl{+}{+}{+}{+} \bigr) \,,
\nonumber \\[0.2em]
  \bigl( \re\ru{\0}{+}{\0}{+} - \re\rl{\0}{+}{\0}{+} \bigr)^2
+ \bigl( \im\ru{\0}{+}{\0}{+} - \im\rl{\0}{+}{\0}{+} \bigr)^2
& \le \ru{\0}{\0}{\0}{\0}\,
    \bigl( \ru{+}{+}{+}{+} - \rl{+}{+}{+}{+} \bigr) \,.
\end{align}
A weaker condition is obtained by taking the sum of these two bounds,
\begin{equation}
  \label{l0+bound}
  \bigl( \re\rl{\0}{+}{\0}{+} \bigr)^2
+ \bigl( \im\rl{\0}{+}{\0}{+} \bigr)^2
\le \ru{\0}{\0}{\0}{\0}\, \ru{+}{+}{+}{+}
  - \bigl( \re\ru{\0}{+}{\0}{+} \bigr)^2
  - \bigl( \im\ru{\0}{+}{\0}{+} \bigr)^2 \,.
\end{equation}
The bounds \eqref{l++bound} and \eqref{l0+bound} have right-hand sides
involving only matrix elements accessible with an unpolarized target
and constrain the matrix elements for longitudinal target polarization
on their left-hand sides.
     
As a second example let us derive conditions which involve only matrix
elements $\ru{}{}{}{}$ and $\rn{}{}{}{}$.  To this end we consider the
matrix 
\begin{equation}
\mathrm{C}' = \half \bigl( \mathrm{C}
            + \mathrm{D}^{\dag}\ms \mathrm{C}\, \mathrm{D} \bigr)
\end{equation}
with
\begin{equation}
\mathrm{D} = \begin{pmatrix}
\;0\; & \;0\; & \;1\; & \;0\; & \;0\; & \;0\; \\
0 & 1 & 0 & 0 & 0 & 0 \\
1 & 0 & 0 & 0 & 0 & 0 \\
0 & 0 & 0 & 0 & 0 & 1 \\
0 & 0 & 0 & 0 & 1 & 0 \\
0 & 0 & 0 & 1 & 0 & 0
\end{pmatrix} \,,
\end{equation}
which is half the sum of the positive semidefinite matrices
$\mathrm{C}$ and $\mathrm{D}^{\dag}\ms \mathrm{C}\, \mathrm{D}$ and
hence positive semidefinite itself.  One readily finds that matrix
elements $\rl{}{}{}{}$ and $\rs{}{}{}{}$ drop out in $\mathrm{C}'$,
which reads
\begin{align}
\mathrm{C}' = \begin{pmatrix}
  \ru{+}{+}{+}{+} & 
  \bigl( \ru{\0}{+}{\0}{+}\bigr)^* &
\; \ru{-}{+}{-}{+} \; &
  \rn{+}{+}{+}{+} &
  - \bigl( \rn{\0}{+}{\0}{+} \bigr)^* &
\; \rn{-}{+}{-}{+} \;
\\[0.4em]
  \ru{\0}{+}{\0}{+} &
  \ru{\0}{\0}{\0}{\0} &
  \ru{\0}{+}{\0}{+} &
  \rn{\0}{+}{\0}{+} &
  \rn{\0}{\0}{\0}{\0} &
  \rn{\0}{+}{\0}{+}
\\[0.4em]
\; \ru{-}{+}{-}{+} \; &
  \bigl( \ru{\0}{+}{\0}{+} \bigr)^* &
  \ru{+}{+}{+}{+} &
\; \rn{-}{+}{-}{+} \; &
  - \bigl( \rn{\0}{+}{\0}{+} \bigr)^* &
  \rn{+}{+}{+}{+}
\\[0.4em]
  -\rn{+}{+}{+}{+} &
  \bigl( \rn{\0}{+}{\0}{+} \bigr)^* &
\; -\rn{-}{+}{-}{+} \; &
  \ru{+}{+}{+}{+} & 
  \bigl( \ru{\0}{+}{\0}{+} \bigr)^* &
\;\ru{-}{+}{-}{+} \;
\\[0.4em]
  -\rn{\0}{+}{\0}{+} &
  -\rn{\0}{\0}{\0}{\0} &
  -\rn{\0}{+}{\0}{+} &
  \ru{\0}{+}{\0}{+} &
  \ru{\0}{\0}{\0}{\0} &
  \ru{\0}{+}{\0}{+}
\\[0.4em]
\; -\rn{-}{+}{-}{+} \; &
  \bigl( \rn{\0}{+}{\0}{+} \bigr)^* &
  -\rn{+}{+}{+}{+} &
\; \ru{-}{+}{-}{+} \; &
  \bigl( \ru{\0}{+}{\0}{+} \bigr)^* &
  \ru{+}{+}{+}{+}
\end{pmatrix} \,.
\end{align}
This matrix has three eigenvalues
\begin{align}
& \ru{+}{+}{+}{+} - \ru{-}{+}{-}{+}
  + \im\rn{+}{+}{+}{+} - \im\rn{-}{+}{-}{+} \,,
\phantom{\frac{1}{2}}
\nonumber \\
& \frac{1}{2} \bigl( 
       \ru{+}{+}{+}{+} + \ru{-}{+}{-}{+}
  + \im\rn{+}{+}{+}{+} + \im\rn{-}{+}{-}{+}
  + \ru{\0}{\0}{\0}{\0} + \im\rn{\0}{\0}{\0}{\0}
  \bigr)
\nonumber \\
& ~\pm \frac{1}{2} \sqrt{
  \bigl( \ru{+}{+}{+}{+} + \ru{-}{+}{-}{+}
    + \im\rn{+}{+}{+}{+} + \im\rn{-}{+}{-}{+}
    - \ru{\0}{\0}{\0}{\0} - \im\rn{\0}{\0}{\0}{\0}
  \bigr)^2
  + 8\, \bigl| \ru{\0}{+}{\0}{+} - i \rn{\0}{+}{\0}{+} \bigr|^2}
\end{align}
and three further eigenvalues obtained by reversing the sign of all
matrix elements $\rn{}{}{}{}$.  Their positivity results in the bounds
\begin{equation}
\bigl( \im\rn{+}{+}{+}{+} - \im\rn{-}{+}{-}{+} \bigr)^2
  \le \bigl( \ru{+}{+}{+}{+} - \ru{-}{+}{-}{+} \bigr)^2
\end{equation}
and
\begin{align}
  \label{un-bound}
 &  2\, \bigl( \re\ru{\0}{+}{\0}{+} + \im\rn{\0}{+}{\0}{+} \bigr)^2
  + 2\, \bigl( \im\ru{\0}{+}{\0}{+} - \re\rn{\0}{+}{\0}{+} \bigr)^2
\nonumber \\
 & \qquad\qquad\qquad \le
\bigl( \ru{\0}{\0}{\0}{\0} + \im\rn{\0}{\0}{\0}{\0} \bigr) \,
\bigl( \ru{+}{+}{+}{+} + \ru{-}{+}{-}{+}
     + \im\rn{+}{+}{+}{+} + \im\rn{-}{+}{-}{+} \bigr) \,,
\nonumber \\[0.2em]
 &  2\, \bigl( \re\ru{\0}{+}{\0}{+} - \im\rn{\0}{+}{\0}{+} \bigr)^2
  + 2\, \bigl( \im\ru{\0}{+}{\0}{+} + \re\rn{\0}{+}{\0}{+} \bigr)^2
\nonumber \\
 & \qquad\qquad\qquad \le
\bigl( \ru{\0}{\0}{\0}{\0} - \im\rn{\0}{\0}{\0}{\0} \bigr) \,
\bigl( \ru{+}{+}{+}{+} + \ru{-}{+}{-}{+}
     - \im\rn{+}{+}{+}{+} - \im\rn{-}{+}{-}{+} \bigr) \,.
\end{align}
Omitting the terms with $\im\ru{\0}{+}{\0}{+}$ and
$\re\rn{\0}{+}{\0}{+}$, one obtains bounds involving only matrix
elements that are accessible with an unpolarized lepton beam.

As we have seen in Section~\ref{sec:cross}, $s$-channel helicity
conserving matrix elements can be extracted from the angular
distribution under the approximation that $s$-channel helicity
changing transitions are suppressed.  The bounds derived in this
section may be used to check the consistency of this approximation.


\section{Mixing between transverse and longitudinal polarization}
\label{sec:mix} 

So far we have discussed target polarization longitudinal or
transverse to the virtual photon direction in the target rest frame,
which is natural from the point of view of the strong-interaction
dynamics.  In an experimental setup one has however definite target
polarization with respect to the lepton beam direction.  The
transformation from one polarization basis to the other is readily
performed using the relations \eqref{S-vs-P}.  For a target having
longitudinal polarization $P_L$ with respect to the lepton beam one
finds
\begin{align}
  \label{target-L}
& \frac{d\sigma}{d\phi\, d\varphi\, d(\cos\vartheta)\,
                 dx_B\, dQ^2\, dt}
 = \frac{1}{2\pi}\, \frac{d\sigma}{dx_B\, dQ^2\, dt}\,
\nonumber \\[0.3em]
&\qquad\qquad \times
   \Bigl( W_{UU} + P_L \Bigl[\ms
                   \cost\, W_{UL} - \sint\, W_{UT}(\phi_S=0) \Bigr]
\nonumber \\[0.3em]
&\qquad\qquad\quad
        + P_\ell\ms W_{LU} + P_\ell\ms P_L \Bigl[\ms
                   \cost\, W_{LL} - \sint\, W_{LT}(\phi_S=0) \Bigr]
   \Bigr) \,.
\end{align}
Note that in this case the azimuthal angle $\psi$ in \eqref{X-sect}
needs to be defined with respect to some fixed spatial direction in
the target rest frame, rather than with respect to the (vanishing)
transverse component of the target polarization relative to the lepton
beam.  We have integrated over this angle in \eqref{target-L} because
the cross section does not depend on~it.

For a target having transverse polarization $P_T$ with respect to the
lepton beam one has
\begin{align}
  \label{target-T}
& \frac{d\sigma}{d\phi_S\, d\phi\, d\varphi\, d(\cos\vartheta)\,
                 dx_B\, dQ^2\, dt}
 = \frac{1}{(2\pi)^2}\, \frac{d\sigma}{dx_B\, dQ^2\, dt}\,
   \frac{\cost}{\rule{0pt}{2.1ex}
                1 - \sin^2\bs\theta_\gamma\, \sin^2\bs\phi_S}\,
\nonumber \\[0.4em]
&\qquad\qquad\qquad \times
   \Biggl( W_{UU} + P_T\, \frac{\cost\, W_{UT}
       + \sint \cos\phi_S\, W_{UL}}{\bigl( \rule{0pt}{2.1ex}
         1 - \sin^2\bs\theta_\gamma\, \sin^2\bs\phi_S \bigr){}^{1/2}}
\nonumber \\[0.3em]
&\qquad\qquad\qquad\quad
   + P_\ell\ms W_{LU} + P_\ell\ms P_T\, \frac{\cost\, W_{LT}
       + \sint \cos\phi_S\, W_{LL}}{\bigl( \rule{0pt}{2.1ex}
         1 - \sin^2\bs\theta_\gamma\, \sin^2\bs\phi_S \bigr){}^{1/2}}
   \Biggr) \,.
\end{align}
The factor $\cost\ms / (1 - \sin^2\bs\theta_\gamma\, \sin^2\bs\phi_S)$
\pagebreak[2]
comes from the change of variables from $d\psi$ to $d\phi_S$ in the
cross section.  The relation between these two angles is readily
obtained by setting $P_L=0$ in \eqref{S-vs-P} and given in
\cite{Diehl:2005pc}.

It is a straightforward (if somewhat lengthy) exercise to insert our
results \eqref{WUL}, \eqref{WLL} and \eqref{WUT}, \eqref{WLT} into
\eqref{target-L} and \eqref{target-T} and to rewrite the expressions
in terms of a suitable basis of functions depending on the azimuthal
angles.  Here we only give the combinations needed in \eqref{target-T}
for a transversely polarized target and an unpolarized beam,
\begin{align}
  \label{LL-ad}
& \hspace{-9em}
\cost\, W_{UT}^{LL}(\phi_S,\phi)
+ \sint \cos\phi_S\, W_{UL}^{LL}(\phi)
  \phantom{\biggl[ \biggr]}
\nonumber \\
= \sin(\phi-\phi_S)\, \biggl[ &
  \cost 
  \im\bigl( \rn{\0}{\0}{+}{+} + \epsilon \rn{\0}{\0}{\0}{\0} \bigr)
  - \sint\ms \sqrt{\epsilon (1+\epsilon)}\ms \im\rl{\0}{\0}{\0}{+}
\nonumber \\
 -& \cos(2\phi)\, \Bigl\{
    \cost\, \epsilon \im\rn{\0}{\0}{-}{+}
    - \sint\ms \sqrt{\epsilon (1+\epsilon)}\ms \im\rl{\0}{\0}{\0}{+}
  \Bigr\}
  \phantom{\biggl[ \biggr]}
\nonumber \\
 -& \, 2\cos\phi\, \Bigl\{
    \cost \sqrt{\epsilon (1+\epsilon)}\ms \im\rn{\0}{\0}{\0}{+}
    + \tfrac{1}{4} \sint\, \epsilon \im\rl{\0}{\0}{-}{+}
  \Bigr\}
\biggr]
\nonumber \\
{}+ \cos(\phi-\phi_S)\, \biggl[ &
  - \sin(2\phi)\, \Bigl\{
    \cost\, \epsilon \im\rs{\0}{\0}{-}{+}
  + \sint\ms \sqrt{\epsilon (1+\epsilon)}\ms \im\rl{\0}{\0}{\0}{+}
  \Bigr\}
\nonumber \\
 -& \, 2\sin\phi\, \Bigl\{
    \cost\ms \sqrt{\epsilon (1+\epsilon)}\ms \im\rs{\0}{\0}{\0}{+}
    + \tfrac{1}{4} \sint\, \epsilon \im\rl{\0}{\0}{-}{+}
  \Bigr\}
\biggr]
\nonumber \\
& \hspace{-7em}{}
- \tfrac{1}{2} \sint\ms \sin(\phi_S + 2\phi)\,
    \epsilon \im\rl{\0}{\0}{-}{+} \,,
  \phantom{\biggl[ \biggr]}
\end{align}
%
\begin{align}
  \label{LT-ad}
& \hspace{-9em}
\cost\, W_{UT}^{LT}(\phi_S,\phi,\varphi)
+ \sint \cos\phi_S\, W_{UL}^{LT}(\phi,\varphi)
  \phantom{\biggl[ \biggr]}
\nonumber \\
= \sin(\phi-\phi_S)\, \biggl[ &
  \cos(\phi + \varphi)\, \Bigl\{
    \cost\ms \sqrt{\epsilon (1+\epsilon)}
    \im\bigl( \rn{\0}{+}{\0}{+} - \rn{-}{\0}{\0}{+} \bigr)
\nonumber \\
& \hspace{4.5em}{}
  + \tfrac{1}{2} \sint\, \Bigl[
    \im\bigl( \rl{\0}{+}{+}{+} - \rl{-}{\0}{+}{+}
            + 2\epsilon\ms \rl{\0}{+}{\0}{\0} \bigr)
    + \epsilon \im\rl{\0}{+}{-}{+} \Bigr]
  \Bigr\}
  \phantom{\biggl[ \biggr]}
\nonumber \\
 -& \cos(\phi - \varphi)\, \Bigl\{
    \cost\ms \sqrt{\epsilon (1+\epsilon)}
      \im\bigl( \rn{\0}{-}{\0}{+} - \rn{+}{\0}{\0}{+} \bigr)
  \phantom{\biggl[ \biggr]}
\nonumber \\
& \hspace{4.5em}{}
    + \tfrac{1}{2} \sint\, \Bigl[
      \im\bigl( \rl{\0}{+}{+}{+} - \rl{-}{\0}{+}{+}
            + 2\epsilon\ms \rl{\0}{+}{\0}{\0} \bigr)
      - \epsilon \im\rl{+}{\0}{-}{+} \Bigr]
  \Bigr\}
  \phantom{\biggl[ \biggr]}
\nonumber \\
 +& \cos(2\phi + \varphi)\, \Bigl\{
    \cost\, \epsilon \im\rn{\0}{+}{-}{+}
    - \tfrac{1}{2} \sint\ms \sqrt{\epsilon (1+\epsilon)}\ms
      \im\bigl( \rl{\0}{+}{\0}{+} - \rl{-}{\0}{\0}{+} \bigr)
  \Bigr\}
  \phantom{\biggl[ \biggr]}
\nonumber \\
 +& \cos(2\phi - \varphi)\, \Bigl\{
    \cost\, \epsilon \im\rn{+}{\0}{-}{+}
    + \tfrac{1}{2} \sint\ms \sqrt{\epsilon (1+\epsilon)}\ms
      \im\bigl( \rl{\0}{-}{\0}{+} - \rl{+}{\0}{\0}{+} \bigr)
  \Bigr\}
  \phantom{\biggl[ \biggr]}
\nonumber \\
 -& \cos\varphi\, \Bigl\{
    \cost\ms \im\bigl( \rn{\0}{+}{+}{+} - \rn{-}{\0}{+}{+}
                     + 2\epsilon \rn{\0}{+}{\0}{\0} \bigr)
  \phantom{\biggl[ \biggr]}
\nonumber \\
& \hspace{2.1em}{}
    - \tfrac{1}{2} \sint\ms \sqrt{\epsilon (1+\epsilon)}\; \Bigl[
      \im\bigl( \rl{\0}{+}{\0}{+} - \rl{-}{\0}{\0}{+} \bigr)
      - \im\bigl( \rl{\0}{-}{\0}{+} - \rl{+}{\0}{\0}{+} \bigr) \Bigr]
  \Bigr\}
\biggr]
\nonumber \\
{}+ \cos(\phi-\phi_S)\, \biggl[ &
  \sin(\phi + \varphi)\, \Bigl\{
    \cost\ms \sqrt{\epsilon (1+\epsilon)}\ms
    \im\bigl( \rs{\0}{+}{\0}{+} - \rs{-}{\0}{\0}{+} \bigr)
\nonumber \\
& \hspace{4.5em}{}
  - \tfrac{1}{2} \sint\, \Bigl[
    \im\bigl( \rl{\0}{+}{+}{+} - \rl{-}{\0}{+}{+}
            + 2\epsilon\ms \rl{\0}{+}{\0}{\0} \bigr)
    - \epsilon \im\rl{\0}{+}{-}{+} \Bigr]
  \Bigr\}
  \phantom{\biggl[ \biggr]}
\displaybreak[2]
\nonumber \\
 -& \sin(\phi - \varphi)\, \Bigl\{
    \cost\ms \sqrt{\epsilon (1+\epsilon)}\ms
      \im\bigl( \rs{\0}{-}{\0}{+} - \rs{+}{\0}{\0}{+} \bigr)
  \phantom{\biggl[ \biggr]}
\nonumber \\
& \hspace{4.5em}{}
    - \tfrac{1}{2} \sint\, \Bigl[
      \im\bigl( \rl{\0}{+}{+}{+} - \rl{-}{\0}{+}{+}
            + 2\epsilon\ms \rl{\0}{+}{\0}{\0} \bigr)
      + \epsilon \im\rl{+}{\0}{-}{+} \Bigr]
  \Bigr\}
  \phantom{\biggl[ \biggr]}
\nonumber \\
 +& \sin(2\phi + \varphi)\, \Bigl\{
    \cost\, \epsilon \im\rs{\0}{+}{-}{+}
    + \tfrac{1}{2} \sint\ms \sqrt{\epsilon (1+\epsilon)}\ms
      \im\bigl( \rl{\0}{+}{\0}{+} - \rl{-}{\0}{\0}{+} \bigr)
  \Bigr\}
  \phantom{\biggl[ \biggr]}
\nonumber \\
 +& \sin(2\phi - \varphi)\, \Bigl\{
    \cost\, \epsilon \im\rs{+}{\0}{-}{+}
    - \tfrac{1}{2} \sint\ms \sqrt{\epsilon (1+\epsilon)}\ms
      \im\bigl( \rl{\0}{-}{\0}{+} - \rl{+}{\0}{\0}{+} \bigr)
  \Bigr\}
  \phantom{\biggl[ \biggr]}
\nonumber \\
 -& \sin\varphi\, \Bigl\{
    \cost \im\bigl( \rs{\0}{+}{+}{+} - \rs{-}{\0}{+}{+}
                 + 2\epsilon \rs{\0}{+}{\0}{\0} \bigr)
  \phantom{\biggl[ \biggr]}
\nonumber \\
& \hspace{2.1em}{}
    - \tfrac{1}{2} \sint\ms \sqrt{\epsilon (1+\epsilon)}\; \Bigl[
      \im\bigl( \rl{\0}{+}{\0}{+} - \rl{-}{\0}{\0}{+} \bigr)
      + \im\bigl( \rl{\0}{-}{\0}{+} - \rl{+}{\0}{\0}{+} \bigr) \Bigr]
  \Bigr\}
\biggr]
\nonumber \\
& \hspace{-7em}{}
+ \tfrac{1}{2} \sint\, \Bigl\{
      \sin(\phi_S + 2\phi + \varphi)\, \epsilon \im\rl{\0}{+}{-}{+}
    + \sin(\phi_S + 2\phi - \varphi)\, \epsilon \im\rl{+}{\0}{-}{+}
  \Bigr\} \,,
  \phantom{\biggl[ \biggr]}
\end{align}
%
\begin{align}
  \label{TT-ad}
& \cost\, W_{UT}^{TT}(\phi_S,\phi,\varphi)
+ \sint \cos\phi_S\, W_{UL}^{TT}(\phi,\varphi)
  \phantom{\biggl[ \biggr]}
\nonumber \\
& \quad = \sin(\phi- \phi_S)\, \biggl[
  \tfrac{1}{2} \cost\ms \im\bigl( \rn{+}{+}{+}{+} + \rn{-}{-}{+}{+}
                                + 2\epsilon \rn{+}{+}{\0}{\0} \bigr)
  - \tfrac{1}{2} \sint\ms \sqrt{\epsilon (1+\epsilon)}\ms
    \im\bigl( \rl{+}{+}{\0}{+} + \rl{-}{-}{\0}{+} \bigr)
  \phantom{\biggl[ \biggr]}
\nonumber \\
 & \hspace{2.5em} - \cos(2\phi)\, \Bigl\{
    \cost\, \epsilon \im\rn{+}{+}{-}{+}
    - \tfrac{1}{2} \sint\ms \sqrt{\epsilon (1+\epsilon)}\ms
      \im\bigl( \rl{+}{+}{\0}{+} + \rl{-}{-}{\0}{+} \bigr)
  \Bigr\}
  \phantom{\biggl[ \biggr]}
\nonumber \\
 & \hspace{2.5em} - \cos\phi\, \Bigl\{
    \cost\ms \sqrt{\epsilon (1+\epsilon)}\ms
      \im\bigl( \rn{+}{+}{\0}{+} + \rn{-}{-}{\0}{+} \bigr)
    + \tfrac{1}{2} \sint\, \epsilon \im\rl{+}{+}{-}{+}
   \Bigr\}
  \phantom{\biggl[ \biggr]}
\nonumber \\
 & \hspace{2.5em} + \tfrac{1}{2} \cos(2\phi + 2\varphi)\, \Bigl\{
    \cost\, \epsilon \im\rn{-}{+}{-}{+}
    - \sint\ms \sqrt{\epsilon (1+\epsilon)}\ms \im\rl{-}{+}{\0}{+}
  \Bigr\}
  \phantom{\biggl[ \biggr]}
\nonumber \\
 & \hspace{2.5em} + \tfrac{1}{2} \cos(2\phi - 2\varphi)\, \Bigl\{
    \cost\, \epsilon \im\rn{+}{-}{-}{+}
    - \sint\ms \sqrt{\epsilon (1+\epsilon)}\ms \im\rl{+}{-}{\0}{+}
  \Bigr\}
  \phantom{\biggl[ \biggr]}
\nonumber \\
 & \hspace{2.5em} - \cos(2\varphi)\, \Bigl\{
    \cost\ms
    \im\bigl( \rn{-}{+}{+}{+} + \epsilon \rn{-}{+}{\0}{\0} \bigr)
    - \tfrac{1}{2} \sint\ms \sqrt{\epsilon (1+\epsilon)}\;
      \Bigl[ \im\rl{-}{+}{\0}{+} + \im\rl{+}{-}{\0}{+} \Bigr]
  \Bigr\}
  \phantom{\biggl[ \biggr]}
\nonumber \\
 & \hspace{2.5em} +  \cos(\phi + 2\varphi)\, \Bigl\{
     \cost\ms \sqrt{\epsilon (1+\epsilon)}\ms \im\rn{-}{+}{\0}{+}
     + \tfrac{1}{4} \sint\, \Bigl[ \epsilon \im\rl{-}{+}{-}{+}
     + 2\im\bigl( \rl{-}{+}{+}{+} + \epsilon\ms \rl{-}{+}{\0}{\0} \bigr)
     \Bigr]
  \Bigr\}
  \phantom{\biggl[ \biggr]}
\nonumber \\
 & \hspace{2.5em} + \cos(\phi - 2\varphi)\, \Bigl\{
    \cost\ms \sqrt{\epsilon (1+\epsilon)}\ms \im\rn{+}{-}{\0}{+}
     + \tfrac{1}{4} \sint\, \Bigl[ \epsilon \im\rl{+}{-}{-}{+}
     - 2\im\bigl( \rl{-}{+}{+}{+} + \epsilon\ms \rl{-}{+}{\0}{\0} \bigr)
     \Bigr]
  \Bigr\}
\biggr]
\nonumber \\
& \quad {}+ \cos(\phi-\phi_S)\, \biggl[
  - \sin(2\phi)\, \Bigl\{
    \cost\, \epsilon \im\rs{+}{+}{-}{+}
    + \tfrac{1}{2} \sint\ms \sqrt{\epsilon (1+\epsilon)}\ms
      \im\bigl( \rl{+}{+}{\0}{+} + \rl{-}{-}{\0}{+} \bigr)
  \Bigr\}
\nonumber \\
 & \hspace{2.5em} - \sin\phi\, \Bigl\{
    \cost\ms \sqrt{\epsilon (1+\epsilon)}\ms 
      \im\bigl( \rs{+}{+}{\0}{+} + \rs{-}{-}{\0}{+} \bigr)
    + \tfrac{1}{2} \sint\, \epsilon \im\rl{+}{+}{-}{+}
  \Bigr\}
  \phantom{\biggl[ \biggr]}
\nonumber \\
 & \hspace{2.5em} + \tfrac{1}{2} \sin(2\phi + 2\varphi)\, \Bigl\{
     \cost\, \epsilon \im\rs{-}{+}{-}{+}
     + \sint\ms \sqrt{\epsilon (1+\epsilon)}\ms \im\rl{-}{+}{\0}{+}
  \Bigr\}
  \phantom{\biggl[ \biggr]}
\nonumber \\
 & \hspace{2.5em} + \tfrac{1}{2} \sin(2\phi - 2\varphi)\, \Bigl\{
    \cost\, \epsilon \im\rs{+}{-}{-}{+}
    + \sint\ms \sqrt{\epsilon (1+\epsilon)}\ms \im\rl{+}{-}{\0}{+}
  \Bigr\}
  \phantom{\biggl[ \biggr]}
\nonumber \\
 & \hspace{2.5em} - \sin(2\varphi)\, \Bigr\{
    \cost\ms
      \im\bigl( \rs{-}{+}{+}{+} + \epsilon \rs{-}{+}{\0}{\0} \bigr)
    - \tfrac{1}{2} \sint\ms \sqrt{\epsilon (1+\epsilon)}\;
      \Bigl[ \im\rl{-}{+}{\0}{+} - \im\rl{+}{-}{\0}{+} \Bigr]
  \Bigr\}
  \phantom{\biggl[ \biggr]}
\displaybreak[2]
\nonumber \\
 & \hspace{2.5em} + \sin(\phi + 2\varphi)\, \Bigl\{
    \cost\ms \sqrt{\epsilon (1+\epsilon)}\ms \im\rs{-}{+}{\0}{+}
    + \tfrac{1}{4} \sint\, \Bigl[ \epsilon \im\rl{-}{+}{-}{+}
    - 2\im\bigl( \rl{-}{+}{+}{+} + \epsilon\ms \rl{-}{+}{\0}{\0} \bigr)
     \Bigr]
  \Bigr\}
  \phantom{\biggl[ \biggr]}
\nonumber \\
 & \hspace{2.5em} + \sin(\phi - 2\varphi)\, \Bigl\{
    \cost\ms \sqrt{\epsilon (1+\epsilon)}\ms \im\rs{+}{-}{\0}{+}
     + \tfrac{1}{4} \sint\, \Bigl[ \epsilon \im\rl{+}{-}{-}{+}
     + 2\im\bigl( \rl{-}{+}{+}{+} + \epsilon\ms \rl{-}{+}{\0}{\0} \bigr)
     \Bigr]
  \Bigr\}
\biggr]
\nonumber \\
& \quad {}+ \tfrac{1}{4} \sint\, \Bigl\{
      \sin(\phi_S + 2\phi + 2\varphi)\, \epsilon \im\rl{-}{+}{-}{+}
    + \sin(\phi_S + 2\phi - 2\varphi)\, \epsilon \im\rl{+}{-}{-}{+}
  \Bigr\}
  \phantom{\biggl[ \biggr]}
\nonumber \\
& \quad {}- \tfrac{1}{2} \sint\ms
  \sin(\phi_S + 2\phi)\,  \epsilon \im\rl{+}{+}{-}{+} \,.
  \phantom{\biggl[ \biggr]}
\end{align}
Compared with \eqref{WUT} and \eqref{WLT} we have changed the order of
terms such that one readily sees which coefficients $\cost
\im\rn{}{}{}{}$ or $\cost \im\rs{}{}{}{}$ receive an admixture from
the same coefficients $\sint \im\rl{}{}{}{}$.  The terms in the last
lines of \eqref{LL-ad} and \eqref{LT-ad} and in the last two lines of
\eqref{TT-ad} involve only coefficients $\sint \im\rl{}{}{}{}$.  They
come with an angular dependence which is absent for $\sint=0$, as is
readily seen by rewriting
\begin{equation}
\sin(\phi_S + 2\phi + m\varphi) =
  - \sin(\phi-\phi_S) \cos(3\phi + m\varphi)
  + \cos(\phi-\phi_S) \sin(3\phi + m\varphi)  \,.
\end{equation}

We see in \eqref{LL-ad} to \eqref{TT-ad} that from the angular
dependence of the cross section for transverse target polarization one
can extract linear combinations of terms $\cost \im\rn{}{}{}{}$ and
$\sint \im\rl{}{}{}{}$ or of $\cost \im\rs{}{}{}{}$ and $\sint
\im\rl{}{}{}{}$.  To separate these terms requires an additional
measurement with longitudinal target polarization.\footnote{%
A corresponding separation for semi-inclusive pion production $ep\to
e\pi X$ has recently been performed in
\protect\cite{Airapetian:2005jc}.}
The expressions \eqref{LL-ad} to \eqref{TT-ad} allow us to see for
which terms the admixture of $\sint \im\rl{}{}{}{}$ terms can be
expected to be small, so that $\im\rn{}{}{}{}$ and $\im\rs{}{}{}{}$
may be determined with reasonable accuracy without such an additional
measurement.  Let us discuss a few examples.
\begin{enumerate}
\item The leading-twist matrix element $\rn{\0}{\0}{\0}{\0}$ appears
  in the linear combination
\begin{align}
c_0 &= \cost 
  \im\bigl( \rn{\0}{\0}{+}{+} + \epsilon \rn{\0}{\0}{\0}{\0} \bigr)
  - \sint\ms \sqrt{\epsilon (1+\epsilon)}\ms \im\rl{\0}{\0}{\0}{+}
\end{align}
in \eqref{LL-ad} and thus has an admixture from $\rl{\0}{\0}{\0}{+}$,
which involves one $s$-channel helicity changing amplitude.  According
to Section \ref{sec:nat-par} this admixture is additionally suppressed
if unnatural parity exchange is small compared with natural parity
exchange.  One may also add to $c_0$ the angular coefficient
\begin{align}
c_1 &= - \cost\, \epsilon \im\rn{\0}{\0}{-}{+}
       + \sint\ms \sqrt{\epsilon (1+\epsilon)}\ms \im\rl{\0}{\0}{\0}{+}
\end{align}
from \eqref{LL-ad}, thus trading the admixture of $\sint\,
\rl{\0}{\0}{\0}{+}$ for an admixture of $\cost\, \rn{\0}{\0}{-}{+}$,
which involves two $s$-channel helicity changing amplitudes (but lacks
the relative factor $\tan\theta_\gamma$ and is not suppressed by
unnatural parity exchange).  We remark that the linear combination of
matrix elements in \eqref{un-n1} corresponds to $c_0 - c_1/\epsilon$,
where $\rl{\0}{\0}{\0}{+}$ does not drop out.  Whether $c_0$, $c_0 +
c_1$ or $c_0 - c_1/\epsilon$ gives the best approximation to $\cost\,
\epsilon \im\rn{\0}{\0}{\0}{\0}$ will thus depend on the detailed
magnitude of the relevant terms.  In practice one might for instance
use the difference between these terms as a measure for the
uncertainty of this approximation.
\item The $s$-channel helicity conserving matrix elements
  $\rn{\0}{+}{\0}{+}$ in \eqref{LT-ad} and $\rn{+}{+}{+}{+}$,
  $\rn{-}{+}{-}{+}$ in \eqref{TT-ad} come together with terms
  involving at least one $s$-channel helicity changing amplitude.
  These admixtures should hence be negligible unless the corresponding
  $s$-channel helicity conserving matrix element is small itself.  For
  $\im\rn{\0}{+}{\0}{+}$ this may for instance happen because of the
  relative phase between the interfering amplitudes.
\item The matrix element $\rn{\0}{\0}{\0}{+}$ in \eqref{LL-ad} comes
  with an admixture from $\rl{\0}{\0}{-}{+}$, which involves two
  $s$-channel helicity changing amplitudes and should hence again be
  suppressed.  In addition, one can extract $\im\rl{\0}{\0}{-}{+}$
  from the angular dependence itself, given the last term in
  \eqref{LL-ad}.  We remark that the unpolarized analog
  $\ru{\0}{\0}{\0}{+}$ of $\rn{\0}{\0}{\0}{+}$ has a real part which
  is experimentally seen to be nonzero
  \cite{Breitweg:1999fm,HERMES-SDME}, providing evidence that
  $s$-channel helicity is not strictly conserved in electroproduction.
  (In the notation of Schilling and Wolf one has $r^5_{00} = -
  \sqrt{2} \re\ru{\0}{\0}{\0}{+}$.)
\item The only $s$-channel helicity conserving matrix elements for
  sideways transverse target polarization in \eqref{LL-ad} to
  \eqref{TT-ad} are $\rs{\0}{+}{\0}{+}$ and $\rs{-}{+}{-}{+}$.  They
  come together with terms involving at least one $s$-channel helicity
  changing amplitude, so that the situation is similar to the one in
  point~2.  Note however that in the present case there is no
  additional suppression of the admixture terms due to unnatural
  parity exchange, since both $\rs{}{}{}{}$ and $\rl{}{}{}{}$ contain
  one unnatural parity exchange amplitude.
\end{enumerate}
In these examples one thus has the favorable situation that the
admixture from longitudinal polarization terms is probably small and
in some cases may even be removed or traded for yet smaller terms.
This does not always happen: the matrix elements $\rn{\0}{+}{-}{+}$
and $\rs{\0}{+}{-}{+}$ in \eqref{LT-ad} receive for instance an
admixture from the $s$-channel helicity conserving term
$\rl{\0}{+}{\0}{+}$, which may not be small itself, so that from the
coefficients of $\sin(\phi-\phi_S) \cos(2\phi+\varphi)$ or
$\cos(\phi-\phi_S) \sin(2\phi+\varphi)$ one cannot directly infer on
the matrix elements $\im\rn{\0}{+}{-}{+}$ or $\im\rs{\0}{+}{-}{+}$.
To make a more precise statement about their size one needs
independent information on $\im\rl{\0}{+}{\0}{+}$, for instance from
the positivity bound \eqref{l0+bound}.


\section{A note on non-resonant contributions}
\label{sec:non} 

So far we have treated the production of two pions in a two-step
picture, where a $\rho$ is first produced in $ep\to ep\rho$ and then
decays as $\rho\to \pi^+\pi^-$.  For deriving the angular distribution
and polarization dependence we have used that the pion pair is in the
$L=1$ partial wave, as can be seen in \eqref{contract-2}.  We did
however not use the narrow-width approximation for the $\rho$ or make
any assumption about its line shape.  In fact, our results for the
angular distribution can readily be used at any given invariant mass
$m_{\pi\pi}$ of the pion pair, with the $ep$ cross sections on the
left- and right-hand sides of \eqref{X-sect} made differential in
$m_{\pi\pi}$.  The spin-density matrix $\rho^{\ms\nu\nu'}_{\mu\mu',
  \lambda\lambda'}$ and its linear combinations $\ru{}{}{}{}$,
$\rl{}{}{}{}$, $\rs{}{}{}{}$, $\rn{}{}{}{}$ then depend on
$m_{\pi\pi}$ and refer not to $\gamma^* p\to \rho p$ but to $\gamma^*
p\to \pi^+\pi^-\, p$ with $\pi^+\pi^-$ in the $L=1$ partial wave.  No
explicit reference to the $\rho$ resonance needs to be made in this
case.

The situation is more complicated if one considers other partial waves
of the pion pair, which can arise from non-resonant production
mechanisms.  To describe a general $\pi^+\pi^-$ state, one should
replace $\rho^{\ms\nu\nu'}_{\mu\mu', \lambda\lambda'}$ with the
spin-density matrix $\rho^{\ms\nu\nu', LL'}_{\mu\mu',
  \lambda\lambda'}$ for a pion pair with angular momentum $L$ in the
amplitude and $L'$ in the conjugate amplitude.  One then has to take
$Y^{}_{L\ms \nu}(\varphi,\vartheta)\, Y^*_{L'\nu'}(\varphi,\vartheta)$
instead of $ Y^{}_{1\ms\nu}(\varphi,\vartheta)\,
Y^*_{1\ms\nu'}(\varphi,\vartheta)$ in \eqref{contract-2} and will
obviously obtain a different angular dependence of the $ep$ cross
section.  The distribution in $\varphi$ and $\vartheta$ for a pion
pair with $L = 0,1,2$ has been discussed in \cite{Sekulin:1973mk}.

It is quite simple to test for the presence of $L=0$ or $L=2$ partial
waves in data by using discrete symmetry properties, and for
$m_{\pi\pi}$ around the $\rho$ mass one can expect that partial waves
with $L=3$ or higher are strongly phase space suppressed.  Since even
partial waves of the $\pi^+\pi^-$ system have charge conjugation
parity $C=+1$ and odd partial waves have $C=-1$, the interference of
$L=1$ with $L=0$ or $L=2$ gives rise to terms in the angular
distribution which are odd under interchange of the $\pi^+$ and
$\pi^-$ momenta, i.e.\ under the replacement
\begin{align}
  \label{C-transf}
\vartheta &\to \pi-\vartheta \,,
&
\varphi &\to \varphi+\pi \,.
\end{align}
Simple examples are an angular dependence like $\cos\vartheta$ or like
an odd polynomial in $\cos\vartheta$.  Corresponding observables
provide a way to study the $L=0$ and $L=2$ partial waves as a
``signal'' interfering with the $\rho$ resonance ``background''
\cite{LehmannDronke:2000xq,Hagler:2002nf}.  This has been used in the
experimental analysis \cite{Airapetian:2004sy}, which did see such
interference away from the $\rho$ resonance peak, whereas close to the
peak the predominance of the $\rho$ was too strong to observe a
significant contribution from any partial wave with $L\neq 1$.  If on
the other hand one is interested in a precise study of the $L=1$
component, one can eliminate its interference with even partial waves
by symmetrizing the angular distribution according to
\eqref{C-transf}.  One is then left with contributions from $L=0$ and
$L=2$ in both the amplitude and its conjugate, which should be very
small around the $\rho$ peak.


\section{Summary}
\label{sec:sum} 

We have expressed the fully differential cross section for exclusive
$\rho$ production on a polarized nucleon in terms of spin density
matrix element for the subprocess $\gamma^* p\to \rho\ms p$.  We work
in the helicity basis for both $\gamma^*$ and $\rho$ and obtain very
similar forms for the unpolarized and polarized parts of the cross
sections, with the substitution rules \eqref{repl-rules-lu} and
\eqref{repl-rules}.  The terms for transverse target polarization
\emph{normal to} the hadron plane closely resemble those for an
unpolarized target, and in both cases the number of independent spin
density matrix elements is reduced if one neglects unnatural parity
exchange compared with natural parity exchange.  The spin density
matrix elements for transverse target polarization \emph{in} the
hadron plane closely resemble those for a longitudinally polarized
target, with both types of matrix elements involving the interference
between natural and unnatural parity exchange.  We have given simple
positivity bounds which involve only matrix elements for an
unpolarized target and either those for longitudinal target
polarization or for transverse target polarization normal to the
hadron plane.  Furthermore, we have investigated the admixture of
longitudinal target polarization relative to the virtual photon
momentum for a target polarized transversely to the lepton beam.  This
admixture should be small for the spin density matrix elements which
conserve $s$-channel helicity in the transition from $\gamma^*$ to
$\rho$, but it may be important for $s$-channel helicity changing
matrix elements.  Finally, we have briefly discussed how the results
obtained in this paper can be used and extended for analyzing the
production of pion pairs not associated with the $\rho$ resonance.


\section*{Acknowledgments}

It is a pleasure to thank my colleagues from HERMES for their interest
in this work and for many discussions, especially A. Borissov,
J. Dreschler, D. Hasch and A. Rostomyan.  I~also gratefully
acknowledge helpful discussions with P. Kroll and A. Sch\"afer.  This
work is supported by the Helmholtz Association, contract number
VH-NG-004.


\end{document}